\documentclass[aps,superscriptaddress,onecolumn,a4paper,11pt,nofootinbib,reprint]{revtex4-2}

\usepackage[colorlinks=true, pdfstartview=FitV, linkcolor=blue, citecolor=red, urlcolor=black]{hyperref}
\usepackage{graphicx}
\usepackage[all]{xy}
\usepackage{amsmath}
\usepackage{amssymb}
\usepackage{tensor}

\usepackage{orcidlink}
\usepackage{amsthm}
\newcommand{\cmmnt}[1]{}
\newcommand{\be}{\begin{equation}}
	\newcommand{\ee}{\end{equation}}
\newcommand{\ben}{\begin{eqnarray}}
	\newcommand{\een}{\end{eqnarray}}
\newcommand{\bes}{\begin{subequations}}
	\newcommand{\ees}{\end{subequations}}
\def\bal#1\eal{\begin{align}#1\end{align}}

\newcommand{\nn}{\nonumber\\}
\newcommand{\bfi}{\begin{figure}}
	\newcommand{\efi}{\end{figure}}
\newcommand{\bc}{\begin{center}}
	\newcommand{\ec}{\end{center}}

\newcommand{\sech}{{\rm sech}}

\newcommand{\p}{{\partial}}
\newcommand{\LL}{{\cal L}}

\newcommand{\Ic}{{\cal I}}

\newcommand{\pu}{\mathrm{\partial_{\mu}}}

\newcommand{\pd}[2]{\ensuremath{\frac{\partial#1}{\partial#2}}}
\newcommand{\deriv}[2]{\ensuremath{\frac{d#1}{d#2}}}
\newcommand{\pb}[1]{\ensuremath{\partial_{#1}}}

\newcommand{\qt}[1]{``#1''}

\begin{document}
	
	\title{Topological solitons of two-field scalar theories in rotationally symmetric backgrounds}
	\author{I. Andrade\,\orcidlink{0000-0002-9790-684X}}
	\email[]{andradesigor0@gmail.com}\affiliation{Departamento de F\'\i sica, Universidade Federal da Para\'\i ba, 58051-970 Jo\~ao Pessoa, PB, Brazil}
	\author{M.A. Liao\,\orcidlink{0000-0001-9720-2079}}
	\email[]{matheusalvesliao@gmail.com}\affiliation{Departamento de F\'\i sica, Universidade Federal da Para\'\i ba, 58051-970 Jo\~ao Pessoa, PB, Brazil}
	\begin{abstract}
	This work concerns scalar field theories with topologically nontrivial vacuum manifold in rotationally symmetric backgrounds of arbitrary dimension. Lagrangians with canonical and generalized kinetic terms are considered, and a radial Bogomol'nyi framework is developed for the symmetric restriction of the theory.  A weak formalism is developed for the second-order radial equations, capable of accommodating the inclusion of weaker singularities introduced by the metric or generalized functions. Next, localized topological solutions are explicitly found. \textcolor{black}{In canonical models, stability of static solitonic solutions  is precluded by scaling arguments, but these concerns can be circumvented with the use of a potential with explicit radial dependence. The stability equations are also derived for general perturbations depending on all coordinates, and their properties are briefly discussed under appropriate regularity and self-adjointness assumptions.} The first-order equations give rise to an integrable orbit equation which can be used to solve the problem completely. It is shown that target space orbits - but not the solutions themselves - are shared between analogous systems defined in different backgrounds. Moreover, the first-order equations can be mapped into a one-dimensional BPS theory through a transformation encoded by a function $\xi(r)$. The internal structure, size and existence of defects follows from the properties and range of this mapping. We use these tools to evaluate the effect of geometry on confinement, existence, and structure of solitons. Exact solutions are provided in Minkowski, Schwarzschild, de Sitter, Schwarzschild de Sitter and conformally flat backgrounds.
	\end{abstract} 
	
	\maketitle
\onecolumngrid	
\section{Introduction}

Real scalar fields are among the most important subjects in classical field theory, both because of their simplicity, which makes them suitable as prototypical examples in investigations of nonlinear equations, and because of their wide range of physical applications~\cite{vachaspati}. In the setting of gravitation and cosmology, scalar fields play important roles in descriptions of inflation~\cite{InflationI, InflationII, InflationIII, InflationIV, InflationV, InflationVI, InflationVII, InflationVIII} and the related post-inflationary preheating~\cite{Preheating, PreheatingII, PreheatingIII, PreheatingIV} and reheating~\cite{reheatingI, reheatingII} phenomena, as well as quintessence~\cite{QuintessenceI, QuintessenceII, QuintessenceIII}, dark matter candidates~\cite{DarkMatterI,DarkMatterII,DarkMatterIII,DarkMatterIV,DarkMatterV, DarkMatterVI, DarkMatterVII,DarkMatterVIII,DarkMatterIX,DarkMatterX}, boson stars~\cite{BosonStar, BosonStarII, BosonStarIII} (and other similar compact astronomical objects), no-hair theorem violations~\cite{NoHairViolationI, NoHairViolationII} among many other applications. Moreover, it is well known that scalar fields can be used to model dynamical degrees of freedom characterizing some alternative theories of gravitation, thus serving both as a tool to probe into the novel features of such theories and as a bridge through which they can be connected to General Relativity (GR) in the appropriate limit~\cite{Chiba,DamourEspositoFarese93,DamourEspositoFarese96, JordanFrameGravity, OlmoI,OlmoII}. In the contexts of supersymmetry and supergravity, scalar fields also play important roles as moduli, and it is common for a single dilaton field to survive at low energy truncations of the theory. Dilatons can also be used to induce supersymmetry breaking. For more insight into these matters, the reader is recommended to~\cite{SupersymmetryI, SupersymmetryII, SupersymmetryIII, SupersymmetryIV, SupersymmetryV, WessZumino1974, Zumino1979, Ketov2009} and references therein.

If the energy functional of a field theory has a topologically nontrivial set of minima, there may exist topological defects, or solitons, in its configuration space. These are finite energy solutions that cannot be smoothly deformed into vacua\footnote[3]{Some references (such as~\cite{rajaraman}) reserve the word \qt{soliton} to objects which, in addition to the requirements above, recover their shapes asymptotically after collision. This is an interesting, but strong requirement usually associated with the integrability of the field equations. Here the word soliton is used in the broader sense common in physics literature (see, for example,~\cite{manton}), so integrability is not a necessary feature.}. One important source of topological solitons is the Kibble-Zurek~\cite{Kibble1976, Zurek} mechanism, which predicts defect formation in systems undergoing spontaneous symmetry breaking. This thermodynamic process can be modeled as a phase transition, and the nontrivial topology follows from the emergence of causally disconnected regions of space, where the order parameter (which can be described as a scalar field) must choose the vacua independently. Cosmological evolution and many important astrophysical processes can be described as phase transitions within rapidly changing thermodynamic conditions, which is why topological solitons have been predicted in such settings for a long time~\cite{Kibble1980,Vilenkin}.

The simplest solitonic configurations are the kinks, which are found in two-dimensional spacetimes and interpolate between two elements of a discrete set of vacua. Kinks can be placed into higher dimensional spacetimes, giving rise to domain walls~\cite{vachaspati, Vilenkin, VilenkinII,Cvetic1996,Cvetic1992}. Ref.~\cite{Cvetic1992} investigates BPS (Bogomol'nyi-Prasad-Sommerfield~\cite{bogo,ps}) domain walls in supergravity, and it is found that gravitational coupling plays a crucial role in the properties and existence of these extended objects, as well as in the nature of supersymmetric vacua. Topological defects in theories with one and two scalar fields are also used as bounce solutions which stabilize the fifth-dimension in Randall–Sundrum theories without recourse to fine-tuning, and to model brane structure (See, for example~\cite{Rubakov1983,Goldberger1999, Kehagias2001, Bazeia2004,BazeiaGomes2004, Dzhunushaliev2006,Dzhunushaliev2008, DeWolfe2000,Gremm2000, Dutra, Dutra2008} and references therein).

The above motivations show that investigations of scalar solitonic configurations in higher-dimensional flat and curved spacetimes are a worthwhile endeavor. Nevertheless, these configurations usually possess infinite energy, for, as is well known, scaling arguments severely restrict finite energy scalar field configurations in more than one spatial dimension. In particular, Derrick's argument~\cite{Derrick} shows that static, finite energy configurations in a canonical scalar theory cannot be stable, as their energy can always be decreased by rescaling.  Nevertheless, the argument is devised for  Minkowski spacetime, and it has been elsewhere shown~\cite{Gonzales, curved,stableDwall} that such scaling arguments can be circumvented for canonical theories in some suitably chosen geometries, although this procedure can be shown to severely restrict the metric, ruling out some important geometries such as Schwarzschild and Reissner-Nordström. The situation is much simpler in generalized models, particularly when the potential is allowed to depend explicitly on the coordinates, in which case stable domain walls can be easily found in any dimension~\cite{PRL2003} , and for a wide variety of spacetimes~\cite{Morris}.  These references manage to evade Derrick's theorem by allowing for a position dependent potential $V(\phi,r)\propto f(r)^{-1}$, where $f(r)$ is a scaling factor specified by the appropriate volume element. This technique can be used to generate stable topological defects in radially symmetric spacetimes of arbitrary dimension. Several subsequent works have built on this idea in order to investigate solitonic scalar configurations in higher-dimensional spacetimes	~\cite{Moreira2022, Moreira2023, Paganelly2021,Paganelly2025, liao2, liao3, Mandal2021, AndradeI, AndradeII, Andrade2019, Casana2015, Casana2020,deSouzaDutra, MoreiradaRocha}, although these mostly deal with Lagrangians with a single scalar field.  We will thus generalize this formalism to the two-field setting and examine the orbital structure of the resulting models. 
	
The object of this paper is the investigation of topological scalar field theories in curved spacetimes, with emphasis on the role of geometry in the existence, asymptotic properties and overall behavior of soliton-like defects. We begin with an overview of BPS field theories depending on a single scalar field. Next, we introduce a new class of two-field Lagrangians which support radially stable topological solutions in any dimension. Models with both canonical and generalized derivative terms of the form introduced in the two-dimensional setting in Ref.~\cite{liao1} are considered. The models with generalized derivative term have been subsequently revisited in other works (see, for example, Refs.~\cite{LCA24,MM,GomesSimas, CasanadaHora,daHora,LimaAlmeida}), especially in the separable case. Extension to higher dimensions is achieved through the technique first used by the authors of Ref.~\cite{PRL2003} in flat spacetime, and later generalized in~\cite{Morris} to general radially symmetric background  geometries. We show that the energy functional, or at least its symmetric restriction, possesses a BPS limit, and the associated orbit equations enforced by Bogomol'nyi saturation are found. It is shown that the orbit structure in a given BPS sector is preserved when the theory is formulated in different spacetimes, and solutions can be (locally) mapped into one another through a transformation parameterized by a single function $\xi(r)$. This realization enables a generalized approach to the problem, which allows us to investigate solution properties without immediate specification of the background geometry. In the next section we provide some simple examples of two-field theories with integrable orbit equation. The first-order equations are solved with the use of the exact orbits and insight into the interplay between field interactions and geometrical effects is gained through investigation of the manner in which defects change when inserted into different spaces. We then conclude with a section devoted to summarize and discuss our results.

\section{One-field theories with BPS saturation in two dimensions}
Let us first consider a theory derived from an action functional of the form:
\begin{equation}\label{ActionI}
	S=\int d^{2}x\sqrt{|g|}\left[\frac{1}{2}g^{\mu\nu}\pu\phi \partial_{\nu}\phi- V(\phi,{x}) \right],
\end{equation}
where $\phi$ is a real-valued scalar field, $g_{\mu\nu}$ is a metric tensor and $g$ is its determinant. If the self-interaction potential is independent of the spatial coordinates, the Lagrangian is said to be of standard, or canonical, form. Throughout this work, we assume that the potential is bounded from below and has a discrete set of degenerate minima, thus allowing for solitonic defects in nontrivial topological sectors of the theory. In Minkowski spacetime, it is well known that energy minimizers are solutions of first-order BPS~\cite{bogo, ps} equations, which saturate an energy bound entirely specified by the boundary data. We start with a line element and self-interaction potential of the form
\begin{equation}\label{ex1}
	ds^2=dt^2-h^2(x)dx^2, \quad \text{and} \quad V(\phi)=\frac{1}{2}W_{\phi}^2
\end{equation}
for some smooth function $h(x)$. Here, $W(\phi)$ is an auxiliary function analogous to a supersymmetric superpotential and $W_{\phi}$ denotes its partial derivative. Note that in this case the introduction of $W(\phi)$ does not incur a significant loss of generality. By following the standard procedure of completing squares in the energy functional~\cite{bogo}, one can easily derive the Bogomol'nyi bound 
\begin{equation}\label{boundI}
	E\geq \left|\int_{\Sigma} W_{\phi}\pd{\phi}{x}\right|dx \equiv \Delta W,
\end{equation}
where $\Sigma$ denotes the spatial domain. If $\Sigma=(-\infty,\infty)$, then $\Delta W= |W(\phi(\infty))-W(\phi(-\infty))|$. Equality is attained if and only if
\begin{equation}\label{BPSI}
\dot{\phi}=0 \quad \text{and} \quad \phi' = h(x)W_{\phi},
\end{equation}
where the dot and prime denote differentiation with respect to time and position respectively. Through  a transformation $d\xi\equiv h(x)dx$, it is possible to write the line element in Eq.~\eqref{ex1} in the form $ds^2=dt^2-d\xi^2$, so the geometry implied by this metric has no intrinsic curvature. Nevertheless, it has other applications outside of traditional gravitational physics. Namely, a nontrivial $h(x)$ may be used to model inhomogeneities of the theory, such as impurities or background fields. In  geometrical optics, a metric tensor of this form can be used to describe electromagnetic waves in spatially varying refractive index in static, nondispersive dielectric media~\cite{Gordon}. Similar geometries can also be used in static, non-flowing, two-dimensional reductions of  analogue gravity theories, where an analogue acoustic line-element $d\tilde{s}^2=c^2(x)\eta^{\mu\nu}dx_{\mu}dx_{\nu}$, conformally related to~\eqref{ex1}, can be obtained in the limit $v\to 0$ (see~\cite{AnalogueGravity} for details).

Background geometries with nonzero scalar curvature are also possible. Since the most general two-dimensional Lorentzian manifold is locally related to a flat space by a conformal transformation, it suffices to consider metric tensors of the form
\begin{equation}
	ds^2=f(x)\left(dt^2-dx^2\right),
\end{equation}
where $f(x)$ is a conformal factor. With this choice of metric, the Euler-Lagrange equations  derived from~\eqref{ActionI} are written
\begin{equation}
	\ddot{\phi} -\phi'' + f(x)V_{\phi}=0.
\end{equation}

If the action is of the form~\eqref{ActionI} with a homogeneous potential $V=V(\phi)$, then the resulting theory does not in general have a BPS limit. Nevertheless, Bogomol'nyi saturation is possible for any $f(x)$ if the canonical model is generalized to allow for a potential of the form
\begin{equation}\label{BPSconformal}
	V(\phi,x)=\frac{1}{2f(x)}W_{\phi}^2,
\end{equation}
in which case the following Bogomol'nyi bound and equations hold 
\begin{equation}\label{BPSII}
	E= \Delta W \iff \phi'(x)= \pm W_{\phi},
\end{equation}	
	where the solution must again be static, as a nonzero kinetic contribution can only increase the energy. By direct differentiation it can be readily verified that BPS solutions automatically solve the Euler-Lagrange equations for this system. Note that the equations are independent of the conformal factor, so that these equations (and the solutions, if the boundary conditions are preserved) are locally identical throughout all reasonable backgrounds. Thus if one, for example, considers the usual $\phi^4$ model specified by $W(\phi)=\phi-\phi^3/3$, then the solution $\phi(x)=\tanh(x)$ holds in the BPS limit for any choice of $f(x)$ in theories generated by the potential~\eqref{BPSconformal}. 

\section{Two-field theories in rotationally symmetric backgrounds}	
	We have  seen that, up to a coordinate redefinition, the BPS equations are locally unchanged by geometry in the two-dimensional coordinate-dependent Lagrangians considered above. Let us see how these results are affected by transition to higher dimensional spacetimes.	We consider a metric representing a static background geometry. Thus, the effect of field configurations on the metric is negligible, a widely used assumption that holds well for fields whose mass is small in the gravitational scale. We may restrict ourselves to line elements of the form
	\begin{equation}\label{metric}
		ds^2=A^2(r)dt^2 - \left[B^2(r)(dr)^2 + \rho^2(r)d\Omega^2_{D}\right],
	\end{equation}  
	where $A, B$ and $\rho$ are smooth nonnegative functions of the radial coordinate and $d\Omega_{D}$ is a differential spanned by the $D-1$ angular variables used to parameterize the $(D-1)$-spheres of the spacetime foliation. We shall exclusively work with coordinate patches such that $A^2$, $B^2$ and $\rho^2$ are real functions. Calculation of the metric determinant leads to
	\begin{equation}\label{density}
		\sqrt{|g|}=|A(r)B(r)\rho(r)^{D-1}|\omega_D(\theta_1,...\theta_{D-2})\equiv\gamma(r)\omega_D,
	\end{equation} 
	where $\omega_D$ is derived from the metric determinant of a unit $(D-1)$-sphere. Polar coordinates correspond to the case $\omega=1$. Extension to cylindrically symmetric geometries amounts to the inclusion of a component $g_{zz}=\zeta^2(r)$, but, from an operational point of view,  this is essentially identical to a redefinition of $\gamma(r)$. One could formally generalize this class of geometries further by making the change $d\Omega_{D}^2\leftrightarrow\omega^{ij}d\theta_{i}\theta_j$, which is also a straightforward generalization as long as this 2-form can be integrated. We thus focus on spherical geometries for notational simplicity, but note that extension to the aforementioned geometries is straightforward.
	
 \cmmnt{As we shall shortly demonstrate in a more general setting that can be reduced to the one-field scenario, the relevant first-order equations in this case are:
	\begin{equation}
		\frac{d\phi}{dr}=\pm \frac{W_{\phi} B^2(r)}{\gamma }
\end{equation}}

	 As is well known, stable, particle-like configurations with finite energy are not possible in canonical models in flat spacetimes with four or more dimensions, while the $D=2$ case is only allowed if $V=0$~\cite{manton}. Nevertheless, generalized models and/or different geometries may be used to evade Derrick's theorem. This may be achieved through the framework introduced in Ref.~\cite{PRL2003}, where it is shown that potentials with explicit radial dependence can be used to generate stable kink-like radial solutions in higher dimensions. Moreover, the potential can be chosen in such a way as to give rise to a \qt{BPS-like} formalism, in the sense that a Bogomol'nyi bound and related first-order equations exist for the symmetric restriction of the theory (meaning that solutions are minimizers at least among configurations possessing radial symmetry). Although the original work pertains to a Minkowski space with polar/spherical coordinates, it is straightforward to generalize the formalism to radially symmetric spacetimes such as those considered in the present work~\cite{Morris}. We now generalize these results to the two-field scenario by writing the action
	 \begin{equation}\label{ActionII}
	 	S=\int d^{D+1}x\sqrt{|g|}\left[\frac{P(\phi,\chi)}{2}g^{\mu\nu}\pu\phi \partial_{\nu}\phi +\frac{Q(\phi,\chi)}{2}g^{\mu\nu}\pu\chi \partial_{\nu}\chi- V(\phi,\chi,r) \right],
	 \end{equation}
	 for smooth functions $P(\phi,\chi)$ and $Q(\phi,\chi)$ which approach constants as the fields tend to a minimum of $V(\phi,\chi,r)$. The field equations derived from the above action functional are:
	  \begin{subequations}\label{EL}
	 	\begin{align}
	 		\nabla_\mu\!\left(P\,\partial^\mu\phi\right)
	 		&=\frac{1}{2}P_{\phi}\,\partial_\mu\phi\,\partial^\mu\phi +\frac{1}{2}Q_{\phi}\,\partial_\mu\chi\,\partial^\mu\chi
	 		-V_{\phi},
	 		\label{ELphi} \\
	 		\nabla_\mu\!\left(Q\,\partial^\mu\chi\right)
	 	&=\frac{1}{2}P_{\chi}\,\partial_\mu\phi\,\partial^\mu\phi +\frac{1}{2}Q_{\chi}\,\partial_\mu\chi\,\partial^\mu\chi
	 	-V_{\chi},
	 		\label{ELchi}
	 	\end{align}
	 \end{subequations}
	  where $\nabla_{\mu}$ is the covariant derivative. A Bogomol'nyi construction is possible in the symmetric case (that is, under the assumption $\phi=\phi(t,r)$) provided that the potential can be written, in terms of a superpotential and the other free functions from the Lagrangian, in the form:

	 \begin{equation}\label{potGen}
	V(\phi,\chi,r)=\frac{B^2(r)}{2\gamma^2(r)}\left(\frac{W_{\phi}^2}{P(\phi,\chi)}+ \frac{W_{\chi}^2}{Q(\phi,\chi)} \right).
	 \end{equation}

The explicit radial dependence on the potential breaks the usual translational invariance in flat spacetimes. Nevertheless, it has been shown that this dependence can appear naturally if the Lagrangian is treated as an effective theory, \textcolor{black}{so that the $r$-dependent potential emerges naturally as the effect of an external electric or scalar field~\cite{PRL2003,Morris,Cvetic1994,Andrade:2023lmn, Casana:2014}. Here we shall closely follow the procedure outlined in Ref.~\cite{Andrade:2023lmn} to show that~\eqref{ActionII} can be viewed as an effective theory derived from the parent Lagrangian:}\be
\LL_{P}=-\frac{\varepsilon(\phi,\chi)}{4}F_{\mu\nu}F^{\mu\nu} +\frac{P(\phi,\chi)}{2}g^{\mu\nu}\partial_\mu\phi\partial_\nu\phi +\frac{Q(\phi,\chi)}{2}g^{\mu\nu}\partial_\mu\chi\partial_\nu\chi -A_\mu j^\mu,
\ee
\noindent where $F_{\mu\nu}\equiv \pu A_{\nu}-\pb{\nu}A_{\mu}$ is the usual Maxwell tensor and $j^{\mu}$ is the current. The effective theory is realized if 
$j^0=-A'(r)/(A^2(r)\gamma)$, $j^i=0$. In the electrostatic regime, Gauss's law is solved by
\begin{equation}
F_{r0}(r)=\frac{1}{A(r)\gamma(r)\epsilon(\phi(r),\chi(r))}.
\end{equation} 

If the generalized permittivity satisfies the constraint 

\begin{equation}
	\frac{1}{\epsilon(\phi,\chi)}=\frac{W_\phi^2}{P(\phi,\chi)}+\frac{W_\chi^2}{Q(\phi,\chi)},
\end{equation}
then the model with potential~\eqref{potGen} is recovered as an effective theory as long as the electric field remains frozen, thus explaining the explicit radial dependence on the Lagrangian of our theory. Moreover, explicit radial dependence on the coordinates can also signal the presence of impurities, which may be used to represent the combined background effect of external field/particles or other inhomogeneities. In scalar field equations, impurities may manifest as a product factor  similar to the ones used here~\cite{Currie:1977zz, Kivshar:1991zz,Fei:1992dk,Fei:1993ni,Kalbermann, KinkKink}. More generally, impurities lead to potentials with explicit coordinate dependence which may give equations similar to~\eqref{FOII} in some cases, both in the one-field~\cite{Andrade:2026xzz, AdamI, AdamII} and in two-field~\cite{2fieldimp, 2fieldimp2} scenarios.


We assume $V(\phi,\chi,r)$ is bounded from below and we may hence take $\text{min} \,\{V(\phi,\chi,r)\}=0$ without loss of generality. If these minima are independent of $r$, they collectively make up the vacuum manifold of the theory, which we define as the set
	\begin{equation}\label{VacuumManifold}
		\mathcal{M}=\left\{(\phi_{0}, \chi_{0}): \forall \, r\in\Sigma, V(\phi_{0}, \chi_{0},r)=0 \right\}.
	\end{equation}

\textcolor{black}{The above definition is valid for \emph{true vacua}, which correspond to points where the energy density has a well-defined value, namely zero. This definition, and thus the meaning of our boundary conditions, must be suitably generalized if $V=0$ is attained only as a limit instead of corresponding to a definite point of field space. This could happen, for example, if $P$ or $Q$ vanish at a critical point. In such cases, which will occur in our two last examples, $V$ merely approaches zero (as long as $W_{\phi}^2/P$ and $W_{\chi}^2/Q$ tend to zero). These cases will be appropriately distinguished below through a rigorous weak limit analysis and regularization procedure, so for the time being let us assume $\mathcal{M}$ corresponds to the true zero set of $V$.} Stability under rescaling (i.e., Derrick's argument) has been thoroughly investigated for theories with potentials of the form~\eqref{potGen} in the one-field scenario. It is straightforward to generalize these results to our theory, since the number of fields does not play an essential role in Derrick's argument. Moreover, we can in fact prove a much stronger stability-related result, namely Bogomol'nyi saturation. Since the line element~\eqref{metric} is independent of $t$, the corresponding background geometry gives rise to a timelike Killing vector $K=\pb{t}$. We may thus define a conserved energy functional which, through a Bogomol'nyi construction~\cite{bogo}, can be written in the form:
	\begin{equation}\label{BPSEnergy}
		E= T + \Omega_D\left\{\int_{\Sigma}\frac{dr\gamma}{2B^2(r)}\left[P\left(\frac{\partial\phi }{\partial r}\mp \frac{W_{\phi} B^2(r)}{P\gamma(r)}\right)^2 + Q\left(\frac{\partial\chi }{\partial r} \mp \frac{W_{\chi} B^2(r)}{Q\gamma(r)}\right)^2\right] + \Delta W \right\}\geq \Omega_D\Delta W,
	\end{equation}
	 where $T$ is the kinetic energy contribution containing the terms quadratic in the time derivatives, $\Omega_D$ is the surface area of a $(D-1)$-sphere, $\Sigma$ is the radial domain, and $\Delta W \equiv \pm \int_{\Sigma} dr \frac{\partial W}{\partial r}= \pm \left.W(\phi,\chi) \right|_{\Sigma}$. 
	 If cylindrical rather than spherical symmetry is assumed, there is one additional integration along the direction of  the symmetry axis, so that the above functional must be seen as the energy per unit length. Equality is attained if and only if the fields are static and solve the first-order equations
		\begin{subequations}\label{FOII}
	\begin{align}
	&\phi'(r)=\pm \frac{W_{\phi} B^2(r)}{\gamma(r) P }\label{BPSphi}, \\
	&\chi'(r)=\pm \frac{W_{\chi} B^2(r)}{\gamma(r) Q }.\label{BPSchi}
\end{align}
		\end{subequations}
	In the above and henceforth, the prime denotes differentiation with respect to $r$. For fixed boundary conditions, solutions of the above first-order equations have the minimum energy possible among configurations with radial symmetry, although solutions with explicit angular dependence must be examined separately (see the stability analysis below). As is usual in soliton theory, we assume that the solution is a vacuum solution outside of a localized region of space \textcolor{black}{(or, in order to be more precise and account for poles or zeroes of $P$ and $Q$, we assume that the potential approaches zero sufficiently fast, in accordance with the regularity conditions discussed below)}, hence $V(\phi, \chi, r)\to 0$ as the boundary is approached. This  requirement amounts to the conditions 
	  \begin{equation}\label{EffectiveBCS}
	 	\left. W_{\phi} \right|_{{\partial\Sigma}}=\left. W_{\chi} \right|_{{\partial\Sigma}}=0,
	 \end{equation}
	 in which $\partial\Sigma$ denotes the boundary and we have assumed that,  \textcolor{black}{either both $P$ and $Q$ attain finite nonzero values at the critical points of $W(\phi,\chi)$, so that elements of $\mathcal{M}$ are valid vacuum solutions of~\eqref{EL}, or, if $P$ or $Q$ vanish at the boundary, that the fields approach the zeroes of $W_{\phi}$ and $W_{\chi}$ sufficiently fast to ensure that the singular ratios $(B^2W_\phi^2)/(2P\gamma^2)$ and $(B^2W_\chi^2)/(2Q\gamma^2)$ vanish as $\partial{\Sigma}$ is approached. A more thorough discussion on regularity and near-boundary behavior will be given shortly.} In some spaces, physical configurations may be restricted to compact sets of the form $\Sigma=[r_{-},r_{+}]$, where the boundary points  correspond to hypersurfaces where $B^2\gamma^{-1}$ has isolated zeroes, which in turn give rise to poles at the field equations and energy density. Fulfillment of~\eqref{EffectiveBCS} is thus necessary for regularization of these poles, as will be seen in some of our examples. If $\left.B^2\gamma^{-1}\right|_{\partial_{\Sigma}}\neq 0$, these boundary conditions cannot be seen as regularization requirements, but they are significant for other reasons, such as ensuring finiteness of the energy in some geometries or matching with a vacuum of the parent system in the effective theory point of view. Moreover, it can be shown that Eqs.~\eqref{FOII} can only give rise to two normalizable zero modes, which are particularly important in a supersymmetric setting, if configurations approach a critical point of $W(\phi,\chi)$ asymptotically. We therefore impose~\eqref{EffectiveBCS} as the boundary conditions of our problem and may thus look for topological solutions in the theory if $\mathcal{M}$ is chosen as a topologically nontrivial manifold. This requirement is automatically achieved if~\eqref{potGen} has a discrete, degenerate set of minima.
	 

A natural issue arises in the generalized models when either $P(\phi, \chi)$ or $Q(\phi,\chi)$ has zeroes or poles, as such points correspond respectively to singularities in the derivative or potential terms. Nevertheless, the solutions, as well as their topological and physical properties, can still be consistently explored in a classical field theory  as long as reasonable regularity conditions are satisfied. In many cases, such interior singularities are removable on-shell, as any poles are compensated by zeros of sufficiently higher order. A more rigorous and general version of these requirements can be achieved through a so-called weak formulation, which we now develop and follow with a brief analysis. For  a rigorous treatment, the reader is recommended to~\cite{WeakSol}. A pair $(\phi, \chi)\in H^1_{\text{loc}} \times H^1_{\text{loc}}$ constitutes a weak solution if, for every pair of smooth test functions $v(r), w(r) \in C_c^\infty$ with compact support in $\Sigma$, the following integral equations hold:
\begin{subequations}\label{weakformulation}
 \begin{align}
	\int_{\Sigma} dr \left[ v'(r) \left( \frac{\gamma}{B^2} P \phi' \right) + v(r) \left( \frac{\gamma}{2B^2} P_{\phi} \phi'^2 + \frac{\gamma}{2B^2} Q_{\phi} \chi'^2 + \gamma V_{\phi} \right) \right] &= 0, \label{eq:weak_2nd_phi}\\
	\int_{\Sigma} dr \left[ w'(r) \left( \frac{\gamma}{B^2} Q \chi' \right) + w(r) \left( \frac{\gamma}{2B^2} P_{\chi} \phi'^2 + \frac{\gamma}{2B^2} Q_{\chi} \chi'^2 + \gamma V_{\chi} \right) \right]  &= 0. \label{eq:weak_2nd_chi}
\end{align}
\end{subequations}


	Derivation of the equations above is straightforward, consisting of multiplication of the static radial equations by $v(r),w(r)$, and then using integration by parts and compact support of the test functions. Roughly speaking, a function belongs to $ H^1_{\text{loc}}$ if it is locally sufficiently smooth for the integral formulation above to make sense. Formally, it means that $\phi$, $\chi$, and their weak derivatives exist and belong to $L^2$ (meaning that the square of their absolute value is Lebesgue integrable)~\cite{WeakSol}. Standard results from Sobolev analysis imply that functions satisfying the above formulation are absolutely continuous on every compact subinterval, so the strong field equations are satisfied almost everywhere, being possibly undefined at isolated points. The need for such an extension when $P$ and $Q$ have zeroes or poles should be clear, as, strictly speaking, the field equations themselves cannot be defined at these points.  Nevertheless, we will see in the later examples of this work that these singularities can be removable or at least regular enough to be integrable. Moreover, even in the simpler model where $P=Q=1$, removable poles may emerge as a consequence of coordinate singularities or degenerate metric points, which may still be treatable within a weak formulation. Thus, Eqs.~\eqref{weakformulation} may be viewed as the \qt{minimal} requirements for our theory, so sufficiently weak singularities may in principle be allowed as long as the above integrals make sense. 	 
	
	Since we are mostly concerned with solutions of the first-order equations, let us assume that Eqs.~\eqref{FOII} are solved almost everywhere, thus allowing a finite, discrete set of isolated singularities. This assumption alone does not suffice to ensure consistency with~\eqref{weakformulation}, so we must additionally assume that the fluxes 
\begin{equation}
	\Pi_\phi(r) \equiv \frac{\gamma(r)}{B^2(r)} P(\phi(r), \chi(r)) \phi'(r)= \sigma W_\phi(\phi(r), \chi(r)),
\end{equation}
are continuous across singularities. An analogous condition must also be imposed for $\chi$, of course. Using~\eqref{BPSII} at either side of a singular point $r_{s}$, one can write the flux conditions: 
\begin{equation}
	\lim_{r\to r_{s}^-}\sigma^{l} W_{\psi}(\phi(r), \chi(r))=	\lim_{r\to r_{s}^+}\sigma^{r} W_{\psi}(\phi(r), \chi(r)),
\end{equation}
where $\psi=\phi, \chi$ and $\sigma^{l,r}=\pm 1$. If the superpotential is at least $C^1$ and the fields are continuous and a single BPS branch is followed (i.e., if the signs at~\eqref{BPSII} are not changed at $r=r_{s}$), the above condition is automatically satisfied. This will be the case of every example dealt with in the  next section, so we have shown that all of these configurations solve the field equations at least in the weak sense. Moreover, this formulation also allows for a wider range of solutions than those considered in this work. For instance, if $r_{s}$ is a critical point of $W(\phi,\chi)$, the signs in Eq.~\eqref{FOII} may change without loss of continuity, although the radial BPS property is lost across the branch change. This could represent a nontrivial defect in the vacuum topological sector, analogous to a kink-antikink pair. By the Sobolev embedding theorem, $H^1_{\text{loc}}$ functions always have an absolutely continuous representative in one-dimensional domains. Thus, the weak formulation developed above ensures the continuity of the fields themselves, while remaining permissible enough to accommodate jump discontinuities in the derivatives or weighted fluxes, although such cases will not be pursued in detail at this time.


The above conditions suffice to ensure that a configuration solves the field equation in a weak sense, even if singularities appear at internal points. However, other than requiring conformity with the prescribed boundary conditions, few restrictions on the behavior of the fields at the boundary are imposed by the weak equations. Thus, finiteness of the energy does not follow automatically. However, this property is ensured if, in addition to the regularity requirements discussed in the previous paragraphs, the energy density (which we define as the integrand of the bulk energy functional) has at most an integrable singularity. Thus, there must exist $\alpha\in (0,1)$ such that the energy density is bounded by  $\left(r-r_{\pm}\right)^{-\alpha}$ within a one-sided neighbourhood of the boundary. If this condition is satisfied by a pair $(\phi, \chi)\in H^1_{\text{loc}} \times H^1_{\text{loc}}$ solving~\eqref{FOII} \emph{with the same sign} throughout the entire domain, then the energy is not only finite, but also consistent with the radially symmetric Bogomol'nyi bound. Thus, Eqs.~\eqref{FOII} are still perfectly valid for all solutions considered in this work, even when $P$ or $Q$ have isolated zeroes as in our two last examples. If the solution changes sign at an interior point of $\Sigma$, the Bogomol'nyi procedure is only locally valid, and the resulting configuration need not be a minimizer among symmetric solutions of this sector. A metric coordinate singularity at the origin or a horizon forces $W_{\phi}/P$ and $W_{\chi}/Q$ to vanish with a sufficiently strong zero at this point. Boundary conditions~\eqref{EffectiveBCS} do not automatically imply finiteness of the energy density (although they are necessary unless $P$ or $Q$ has poles, which shall not be the case for any of the models considered here). Nevertheless, we will see that these conditions are needed to soften the singularities and allow for integrability. For instance, regularity of the energy for solutions of~\eqref{FOII} near a boundary point $r_{-}$ requires  $\int_{r_-}^{\delta} dr \left(\phi'W_{\phi}+\chi'W_{\chi}\right)$ for some $\delta >0$. Thus, the energy density remains finite over any compact interval of the origin as long as the products $\phi'W_{\phi}$ and $\chi'W_{\chi}$ tend to zero  faster than $(r-r_{-})^{-1}$. Since any physical measurement of energy must be performed within a finite interval, such weak divergences do not preclude the  viability  of  solutions at the classical level. In all models presented in this work, we shall check the boundary behavior in order to explicitly verify finiteness of the energy.

Another issue associated with zeros or poles of $P$ and $Q$ is the fact that the definition of vacua according to~\eqref{VacuumManifold} is not strictly valid if, e.g., $W_{\phi}$ and $P$ vanish at the same target space point. In this case, the solution and boundary conditions are still valid in our weak formulation as long as $(\phi,\chi)\to (\phi_{0},\chi_{0})$ asymptotically in such a way that $W_{\phi}^2(\phi,\chi)/P(\phi,\chi)\to 0$. The main remaining subtlety lies in the interpretation of the boundary values $(\phi_{0},\chi_{0})$. Indeed, the equations are undefined at this point, so no homogeneous solution coinciding with these exists, and thus they cannot be identified with vacua, despite their role in the determination of boundary conditions and topology. Pragmatically, two choices remain. One possibility is to view such models as a vacuumless analogue of the other systems considered within the work, treating the existence of weak solutions as the fundamental result at the classical level. Alternatively, one may attempt to devise a regularization scheme such that the source of the singularity, say a zero of $P(\phi,\chi)$, is seen as the $\epsilon\to 0$ limit of a family of theories defined by $P_{\epsilon}=P+\epsilon$, where $\epsilon>0$. This should work provided the solution is patchwise regular and the weak limit is well defined on every compact interval containing the singularity. This regularization scheme can be made rigorous with the use of results of Sobolev and Picard–Lindelöf theory, but for our purposes it will be sufficient to note that this regularization is possible for reasonable systems. Pragmatically, this can be verified numerically by solving the differential equations with gradually smaller values of $\epsilon$ (without change of boundary data) and comparing the ensuing solutions. Thus, these theories can be seen as the limit of a family of well-behaved models, each of which has well-defined vacua consistent with the critical points of $W(\phi,\chi)$. The physical interpretation of the boundary conditions in the singular model can be obtained from the regular ones through the $\epsilon\to 0$ limit. \textcolor{black}{Accordingly, in the singular $\epsilon=0$ examples that will be considered in Sec.~\ref{sec:examples}, we shall refer to such endpoints as limiting boundary values or zero-energy limits of the weak problem, reserving the term true vacuum for nonsingular or $\epsilon$-regularized theories.}

	 To find explicit solutions, we seek the orbits described by fields in target space~\cite{RajaramanI, OrbitsFirstorder}. This amounts to derivation of an expression of the form $F(\phi,\chi)=C$ which must be satisfied by all configurations consistent with the first-order equations. Here,  $C$ is a constant. If there exists, as is often the case, a continuous interval of allowed values of $C$ which are compatible with a given set of boundary conditions, then $C$ signals a zero mode of~\eqref{FOII}, since variation of this constant leads to a continuum of solutions with the same energy. To find the orbits explicitly, we multiply the first equation in~\eqref{FOII} by $dx/d\chi$ and write	
			\begin{equation}\label{orbiteq}
			\frac{d\phi}{d\chi}=\frac{W_{\phi}Q}{W_{\chi}P}\implies W_{\chi}P d\phi-W_{\phi}Qd\chi=0 
	\end{equation}
which is the orbit equation. The above equation can be solved exactly - thus giving the general orbit - provided that there exists an integrating factor that turns the above expression into a perfect differential. By definition, this requirement is equivalent to the equations:
\begin{equation}\label{orbitequations}
\frac{\p F}{\p\phi} = \Ic \frac{W_\chi}{Q}  \quad \text{and} \quad \frac{\p F}{\p\chi} = -\frac{\Ic W_\phi}{P},
\end{equation}
	where $\Ic=\Ic(\phi,\chi)$ is an integrating factor. By differentiating the above equations and demanding equality of mixed second derivatives of $F(\phi,\chi)$, we can derive a constraint equation for the orbit. Since the equations are nonlinear, there is no universally applicable method to derive $\mathcal{I}$, but it is often possible to find integrating factors by using an ansatz obtained by simple inspection or educated guesses and then demanding consistency with the exactness constraint.                

We note that, although the first-order equations display an explicit dependence on the radial coordinate and general metric structure of the model, this dependence completely vanishes from the orbit equations.  This means that orbits are preserved by the change of geometry, despite the fact that the solutions are, of course, geometry-dependent. This is a very interesting result, as it highlights a connection between configurations defined in different geometries and solving distinct sets of Euler-Lagrange equations. We note that this relationship follows from Bogomol'nyi saturation, as the geometric factors cannot in general be eliminated from~\eqref{EL} unless the first-order equations are satisfied. As was the case in $D=1$, the relationship between solutions of the first-order equations in different geometries can be succinctly captured by a variable $\xi$, namely,
\begin{equation}\label{transform}
	\frac{d\xi(r)}{dr}=\frac{B^2(r)}{\gamma(r)}\implies \xi(r)=\int dr \frac{B^2(r)}{\gamma(r)} + \xi_{0},
\end{equation}  
in which $\xi_{0}$ is a constant that must conform to the boundary conditions of the problem as well as the orbit. There often exists an interval such that every $\xi_{0}\in I$ leads to an acceptable solution within a given topological sector, in which case this parameter signals the presence of a zero mode of the first-order equations. In terms of $\xi(r)$, Eqs.~\eqref{FOII} are written:
	\begin{align}\label{FOxi}
		\frac{d\phi}{d\xi}=\pm \frac{W_{\phi}}{P}, \quad \quad \quad  \quad \quad \quad  
		\frac{d\chi}{d\xi}=\pm \frac{W_{\chi}}{Q}.
	\end{align}

 These equations lead directly to the orbit equation~\eqref{orbiteq} and can be used to find solutions $(\phi(\xi),\chi(\xi))$ valid for any geometry, at least in a neighbourhood where~\eqref{transform} is well-defined. The above equations have the same form as those of a canonical (if $P=Q=1$), or generalized according to Ref.~\cite{liao1}, $D=1$ system with the BPS property. By way of~\eqref{EffectiveBCS}, the boundary conditions from the original problem can also be mapped into the higher-dimensional setting. Thus, solutions of the first-order equations in a background defined by~\eqref{metric} can be obtained by simply subjecting a solution of $(\phi(x),\chi(x))$ to the transformation $x\leftrightarrow \xi$ with the use of~\eqref{transform}. These features will be more thoroughly discussed in the next section with some explicit examples.

We finish this section with a brief discussion on the linear stability of our solutions. As explained above, the Bogomol'nyi procedure developed above ensures minimization only among radially symmetric configurations satisfying the prescribed set of boundary conditions. Since such solutions share the symmetry of backgrounds compatible with~\eqref{metric}, this is a reasonable assumption, but it is nevertheless natural to inquire whether energetically favorable states may be found once the assumption of symmetry is dropped and nontrivial angular dependence is thus allowed. \textcolor{black}{In what follows, we assume $P$ and $Q$ are strictly positive almost everywhere and that perturbations preserve boundary conditions, produce no extra boundary terms, and have finite weighted norm. $P$ or $Q$ are allowed to vanish at isolated points, provided the corresponding singularities in the fluctuation equations and in the second variation of the energy are integrable and produce no additional boundary terms under integration by parts. Such zeroes form a set of measure zero, and thus do not by themselves make the weighted norm degenerate for finite-energy perturbations. If the zero set of $P$ or $Q$ has nonzero measure, the weighted norm degenerates. The simple argument below does not account for such cases, and we do not claim validity of the following stability argument for models violating any of the assumptions mentioned in this paragraph.} 
 	
We now derive the linear stability equations of the theory by introducing the first-order perturbations $(\phi(r),\chi(r))\to (\phi(r)+\delta\phi(t,\mathbf{x}),\chi(r)+\delta\chi(t,\mathbf{x}))$ and linearizing the field equations~\eqref{EL}. The ensuing equations are separable, so that it shall prove sufficient to consider perturbations of the form $\phi(r)\to\phi(r)+\eta(r)Y(\boldsymbol{\theta})\cos(\omega t)$ and $\chi(r)\to\chi(r)+\zeta(r)Y(\boldsymbol{\theta})\cos(\omega t)$, where $\boldsymbol{\theta}$ represents the $D-1$ angular coordinates. In spherically symmetric spacetimes of arbitrary dimension, the angular equation is solved by hyperspherical harmonics, which are the eigenfunctions of the Laplace–Beltrami operator on the unit sphere $S^{D-1}$~\cite{Hyperspherical}. As is well known, the eigenvalues of this operator are given by $\Lambda_{\ell}=\ell\left(\ell + D -2\right)$, where $\ell$ is a nonnegative integer. Substitution of this result into the separated radial equation leads to

 \bes\label{Eq23}
 \bal
 &-\frac{1}{\gamma}\left(\frac{\gamma P \eta'}{B^2}\right)' -\frac{1}{\gamma}\left[\frac{\gamma}{B^2}\left(P_\phi \phi' \eta + P_\chi \chi'\zeta\right)\right]' -\frac{1}{B^2}\bigg[P_\phi \phi' \eta' +\frac{1}{2}P_{\phi\phi}\phi'^2\eta +\frac{1}{2}P_{\phi\chi}\phi'^2\zeta' \nn
 &+Q_\phi \chi'\zeta +\frac{1}{2}Q_{\phi\phi}\chi'^2\eta +\frac{1}{2}Q_{\phi\chi}\chi'^2\zeta\bigg] +V_{\phi\phi}\eta +V_{\phi\chi}\zeta +\frac{\ell(\ell+D-2)}{\rho^2}\eta= \frac{\omega^2 P \eta}{A^2}\\  \text{and} & \nonumber\\
 &-\frac{1}{\gamma}\left(\frac{\gamma Q \zeta'}{B^2}\right)' -\frac{1}{\gamma}\left[\frac{\gamma}{B^2}\left(Q_\chi \chi' \zeta + Q_\phi \phi'\eta\right)\right]' -\frac{1}{B^2}\bigg[Q_\chi \chi' \zeta' +\frac{1}{2}Q_{\chi\chi}\chi'^2\zeta +\frac{1}{2}Q_{\chi\phi}\chi'^2\eta' \nn
 &+P_\chi \phi'\eta +\frac{1}{2}P_{\chi\chi}\phi'^2\zeta +\frac{1}{2}P_{\chi\phi}\phi'^2\eta\bigg] +V_{\chi\chi}\zeta +V_{\chi\phi}\eta +\frac{\ell(\ell+D-2)}{\rho^2}\zeta= \frac{\omega^2 Q \zeta}{A^2}.
 \eal
 \ees

 A complete spectral analysis lies beyond the scope of this paper, but the linear stability of the solution can be shown, under reasonable assumptions, with a simple argument. Indeed, it is found that the $\ell=0$ case corresponds to radially symmetric fluctuations. This observation conforms to physical intuition, as the rotational energy associated with modes with nonzero angular momenta is expected to increase the eigenvalues. A rigorous demonstration of this assertion  follows from linearization of the radially symmetric field equations  - obtained from substitution of $(\phi,\chi)=(\phi(r),\chi(r))$ into Eqs~\eqref{EL} - from which an equation identical to the $\ell=0$ case of~\eqref{Eq23} is obtained. \textcolor{black}{Under our stated  assumptions, the quadratic form associated to the second variation of the energy for the $\ell$-th angular sector differs from the radial one only by a nonnegative integral. For the singular examples with isolated zeroes of $P$ or $Q$, this statement is to be understood either for the $\epsilon$-regularized theories or for singular limits in which the corresponding quadratic forms converge without extra boundary contributions.} Stability of radially symmetric configurations in similar models has been demonstrated in several references (see, for example,~\cite{PRL2003, Morris}), and the Bogomol'nyi formalism developed above does ensure stability of radially symmetric configurations for given boundary conditions. From these considerations, we deduce that the stability spectrum corresponding to radial fluctuations contains no negative eigenvalues. \textcolor{black}{Moreover, since $\ell\geq 0$ and $D\geq 2$, the $\ell$-dependent contribution is nonnegative and cannot give rise to a negative direction in the admissible perturbation space. Thus, under the regularity and self-adjointness assumptions stated above, angular perturbations do not destabilize solutions of~\eqref{FOII}.}
\section{Examples}\label{sec:examples}
\subsection{Models with canonical derivative terms ($P=Q=1$)}\label{subsec:canonical}
To exemplify the framework discussed above, let us work with the auxiliary function
	\be\label{SPBNRT}
W(\phi,\chi) = \phi -\frac13\phi^3 -a\phi\chi^2,
\ee
where $a$ is a positive parameter of the model. This superpotential was introduced in Ref.~\cite{bnrt1} for the canonical $D=1$ model, and further investigated in subsequent works (see, for example, ~\cite{bnrt2,fator1,BNRT3, BNRT4}), and the theory derived from it is referred to as BNRT (Bazeia, Nascimento, Ribeiro, Toledo) in the literature. The vacuum manifold of the theory is the set $\mathcal{M}=\{(\pm 1, 0), (0,\pm 1/\sqrt{a})\}$. In this case, the first-order equations are
\begin{equation}\label{FOBNRT}
\phi^\prime = \pm\left(1-\phi^2-a\chi^2\right)\frac{B^2(r)}{\gamma(r)},\quad\quad  \quad \quad
\chi^\prime = \mp2a\phi\chi\frac{B^2(r)}{\gamma(r)}.
\end{equation}

\cmmnt{
Direct substitution of $W$ in~\eqref{vinculoI} leads to 
\be\label{intfactdeduction}
\Ic_{\chi}\chi+\Ic=\frac{\partial}{\partial\chi}\left(\chi \Ic(\phi,\chi)\right)=-\frac{1}{aP(\chi)}\left(\Ic+2a^2\Ic_{\phi}(\phi,\chi)\right).
\ee
}

 The authors of reference~\cite{fator1} have shown that the factor $\Ic=\chi^{-(a+1)/a}$ may be used to integrate the orbit equation completely in the original model. As explained above, this integrating factor is still valid in other geometries, and we may thus use the following orbit
\begin{equation}\label{orbits}
	\frac{a\chi^2}{2a-1} + \beta\chi^{\frac{1}{a}}  + 1 =\phi^2,
\end{equation}
where $\beta$ is a constant. We can thus solve for $\phi$ to find two branches, namely $	\phi^{\sigma}(\chi)=\sigma
\sqrt{1+a\chi^2/\left(2a-1\right)+\beta\chi^{1/a}},$ where $\sigma=\pm 1$. Substitution of these orbits into the second equation of~\eqref{FOBNRT} leads to
\begin{equation}
	\int^{\chi(\xi)}
	\frac{d\tilde\chi}
	{\tilde\chi
		\sqrt{
			1+\frac{a}{2a-1}\tilde\chi^2
			+\beta\tilde\chi^{1/a}
	}}
	=
	-2a\sigma\,\xi + C
\end{equation}
for some constant $C$. Different choices of $\beta$ may be used to deform the orbit in such a way as to ensure that different vacua are crossed by the orbit, and fixing this parameter suffices to solve the problem if a solution exists in the topological sector.  The above integral leads to solutions for arbitrary values of $a$ and $\beta$, but some convenient choices of these constants may lead to simpler closed-form expressions. For example, the choice $\beta=0$ gives $\phi^2 + a\chi^2/(1-2a) =1$,
which corresponds to an elliptical orbit if $a\in (0,1/2)$. The orbit passes through the minima $(\phi,\chi)=(\pm1, 0)$ and may thus be used for the sector $(-1,0)\to (1,0)$ as well as its $Z_2$ conjugate. We can use this orbit to integrate the (upper signs) first-order equations and find the solution
\begin{equation}\label{solutionI}
	\phi(\xi)= \tanh(2a\xi),  \quad \quad \quad \quad \quad 	\chi(\xi)=\sqrt{\frac{1-2a}{a}}\,\sech(2a\xi),
\end{equation}
which is fully specified after a line-element is given to determine $\xi(r)$. The simplest explicit examples are found in flat geometries, which are achieved by identification of $\Omega_{D}$ with the surface of a $(D-1)$-sphere. In this case, $\gamma=r^{D-1}$, $B=1$ and thus~\eqref{transform} leads to
\begin{equation}\label{xiflat}
	\xi(r)=\begin{cases}
&\ln(r) + \xi_{0}  \quad \text{if $D=2$},\\
&-\frac{r^{2-D}}{D-2}  + \xi_{0} \quad \text{if $D\geq 3$}.
	\end{cases}
\end{equation}
 
\begin{figure}[t!]
	\centering
	\includegraphics[width=0.48\linewidth]{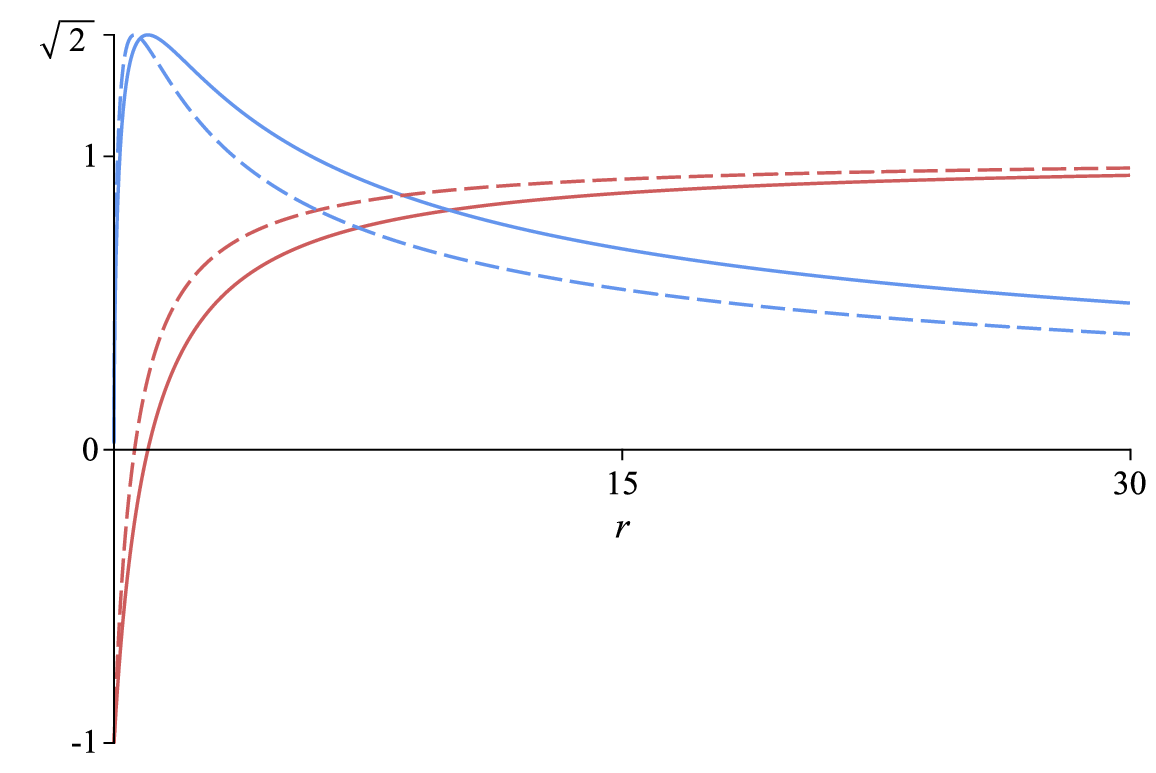}
	\includegraphics[width=0.48\linewidth]{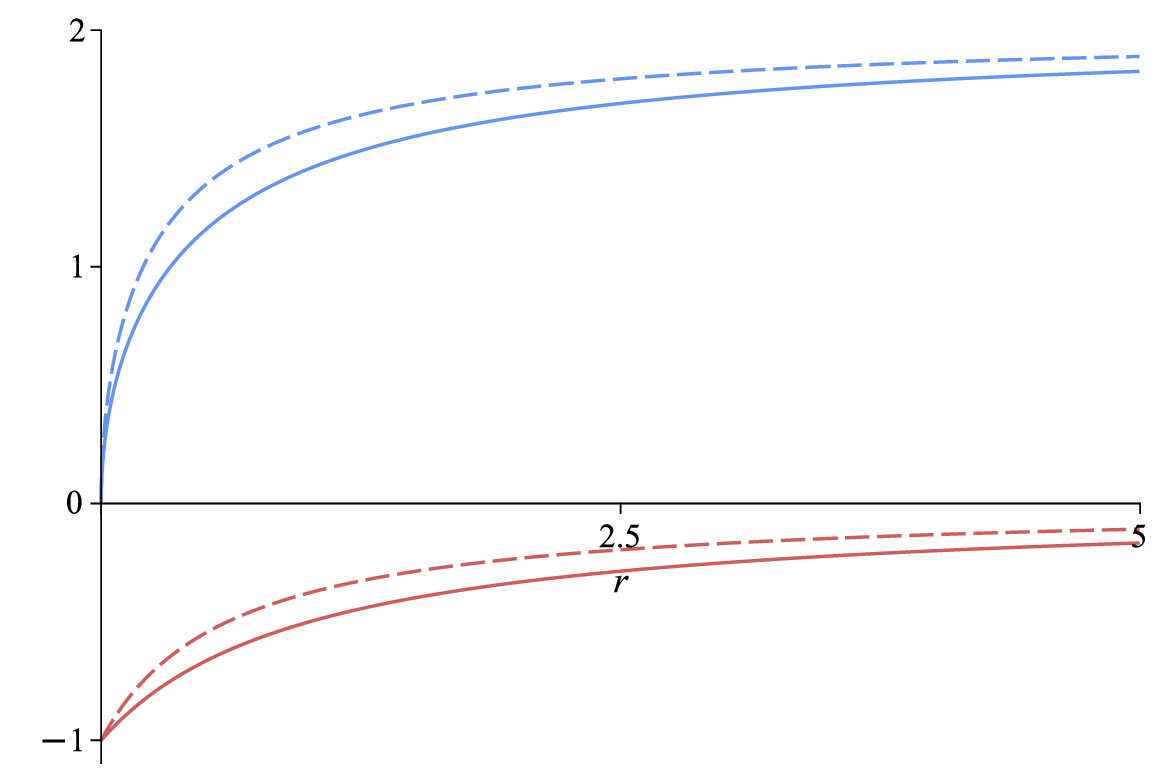}
	\caption{Solutions~\eqref{Eq28} (left) and~\eqref{Eq29} (right) with  $a=1/4$. Here, $\phi(r)$ and $\chi(r)$ are depicted respectively as red and blue plots. Solid and dashed lines correspond respectively $\xi_{0}=0$ and  $\xi_{0}=1/2$.}
	\label{fig1}
\end{figure}

 Substitution of the functions above into~\eqref{solutionI} gives solutions of~\eqref{BPSII} for the given geometry. Let us first consider the problem in flat $D=2$ spacetime. This case closely resembles the $D=1$ theory in that $\xi$ ranges from $-\infty$ to $\infty$ just as the original $x$ variable. At the origin $\gamma$ vanishes due to a coordinate singularity in the polar representation, and regularization implies $W_{\phi}(r=0)=W_{\chi}(r=0)=0$. At infinity, these functions must also vanish in order to avoid logarithmic divergences in the energy functional. The boundary conditions thus map $\xi=\pm\infty$ into the minima $(\phi,\chi)=(\pm 1, 0)$, with arbitrary choices of $\xi_{0}$ representing energetically equivalent solutions of the problem. We may use log and hyperbolic relations to write the solution in the simplified form:
\begin{equation}\label{Eq28}
	\phi(r) =\frac{(r/r_0)^{4a} - 1}{(r/r_0)^{4a} + 1}, \quad \quad \quad \quad \quad \chi(r) = \sqrt{\frac{1-2a}{a}} \frac{2(r/r_0)^{2a}}{(r/r_0)^{4a} + 1},
\end{equation}
where $r_{0}\equiv e^{-\xi_{0}}$. This solution belongs to the sector  $(-1,0)\to (1,0)$. This is true for any choice of $\xi_{0}$, as changes in this parameter function as a rescaling of the radial coordinate. This arbitrary constant signals the presence of a zero mode in the first-order equations, as it parameterizes a family of solutions with the same energy within this topological sector. Such modes are related to symmetries of the first-order equations. In the original model, one encounters an analogous zero mode due to invariance of the equations under the translation $x\to x+x_{0}$, while in $D=2$ the first-order equations can be verified to be invariant under the scale transformation $r\to r/r_{0}$. These solutions are depicted as the left-hand side in Fig.~\ref{fig1} for the case $a=1/4$ with two choices of $\xi_{0}$.

As in the original model, it is also possible to find closed-form solutions in other sectors of the theory~\cite{fator1}. One interesting possibility arises when $a=1/4$, in which case it can be verified that the value $\beta=1/16$ turns the left-hand side of~\eqref{orbits} into a perfect square, from which one may deduce the orbit $\chi=\pm 2\sqrt{1\pm\phi}$. We thus have four distinct branches, which can be used to find solutions in different topological sectors. Choosing the $(+,+)$ branch we find the solution:
\begin{equation}\label{Eq29}
	\phi(r) = - \frac{r_{0}}{r + r_{0}}, \quad \quad \quad \quad \quad \chi(r)=2 \sqrt{\frac{r}{r + r_{0}}}.
\end{equation}
This solution is shown as the right-hand plot of Fig.~\ref{fig1}. Note that the derivatives do not vanish at $r=0$. Indeed, a Taylor expansion about the origin shows that $\phi'(0)=1/r_{0}$ to dominant order, while $\chi'$ diverges as $1/\sqrt{r}$ in the same neighbourhood. The divergence in $\chi'^2/2$ is however precisely compensated by the $\gamma(r)=r$ factor from the Jacobian, thus giving rise to a perfectly smooth energy density because of the boundary conditions~\eqref{EffectiveBCS}. \textcolor{black}{For the solution corresponding to~\eqref{Eq28}, the situation is slightly different. The field derivatives behave as $\phi\sim r^{4a-1}$, $\chi\sim r^{2a-1}$. Since $a\in(0,1/2)$ for this solution, the energy density (divided by $\Omega_D$) is dominated by the latter derivative near the origin, where it behaves as $\mathcal{H}= r\left(\phi'^2+\chi'^2\right)\sim r^{4a-1}$. Since $a>0$ by hypothesis, $\int_{0}^{\delta} dr \mathcal{H}$ is bounded for all $\delta$, so the energy is always finite and in fact equal to the Bogomol'nyi bound. Moreover, the fields and their derivatives are also integrable within the entire domain, so these quantities are always physically meaningful within a neighbourhood of nonzero size, which is sufficient at the classical level this paper is concerned with. }

Interestingly, the situation is substantially different in higher dimensions. If $\xi=-1/r + \xi_{0}$ is substituted into~\eqref{solutionI}, one notices that the vacua are never reached for any finite choice of $\xi_{0}$. This happens because, for $D\geq 3$, $\xi$ changes from $-\infty$ to $\xi_{0}$ as $r$ grows from $0$ to infinity, while the vacua of this model can only be attained for $\xi\to\pm\infty$. This can be proved by expanding ~\eqref{FOBNRT} in a left neighbourhood of $\xi=\xi_0$. The boundary conditions thus impose $\phi'(\xi_0)=\chi'(\xi_0)=0$. The first-order correction can be obtained by differentiating~\eqref{FOBNRT} to find $\phi''(\xi_0)=-2\left(\phi(\xi_0)\phi'(\xi_0) +a\chi(\xi_0)\chi'(\xi_0) \right)=0$ and $\chi''(\xi_0)=-2a\left(\chi(\xi_0)\phi'(\xi_0)+\phi(\xi_0)\chi'(\xi_0)\right)=0$. By repeating this process, it can be shown by induction that all derivatives of $\phi$ and $\chi$ vanish under this assumption, and thus the configuration can reach a point of $\mathcal{M}$ at a finite value of $\xi$ only if the solution is constant.

 Thus, we see that nontrivial solutions satisfying~\eqref{EffectiveBCS} are impossible in flat spacetimes with $D\geq 3$. Moreover, the above demonstration is also unchanged for any curved spacetime such that $\xi\in (-\infty,\xi_{0})$. Such geometries include, for example, the exterior Schwarzschild background, whose line element is
\begin{equation}\label{Sch}
		ds^2=\left(1-\frac{r_s}{r}\right)dt^2 - \left(1-\frac{r_s}{r}\right)^{-1}dr^2+ r^2d\Omega^2 \implies \, \xi(r)=\frac{\ln(r-r_{s})-\ln(r)}{r_s}
\end{equation}
where $r_s$ is the Schwarzschild radius and we are considering the exterior region $r> r_s$, so that $A^2(r)=B^{-2}(r)=\left(1-r_s/r\right)$ remains nonnegative everywhere. Since $\xi$ approaches $-\infty$ and $\xi_{0}$ when $r$ tends respectively to $r_s$ and infinity, the range of $\xi$ is the same as in flat $D=3$ spacetime, thus precluding topological solutions. Configurations valid in a Schwarzschild background will be found in subsection~\ref{subsec:noncanonical} with use of Lagrangians compatible with compact solutions.

The above issue is clearly encountered in any asymptotically flat spacetime. To find configurations conforming to~\eqref{EffectiveBCS} in $D\geq 3$, we must therefore change either the vacuum manifold of the theory or the large $r$ behavior of the metric tensor. The latter approach can be effected by considering any geometry such that $\xi(r_{min})=-\infty$ and $\xi(r_{max})=\infty$, where $r_{min}$ and $r_{max}$ denote the boundary of the domain. In this case, the first-order equations become closely analogous to those of a $D=1$ BPS field theory in flat space, where the same $x\in(-\infty,\infty)$ range is encountered. These requirements force us to consider spacetimes which are not asymptotically flat.\footnote[2]{ As is well known, an unambiguous definition of energy in the style of the Arnowitt–Deser–Misner (ADM) formalism~\cite{ADM} is not possible in such spaces. If the metric is of the form~\eqref{metric} within a given patch, a timelike Killing vector $K=C\partial_{t}$ always exists,
so the conserved current $J^{\mu}=K^{\nu}T^{\mu}_{\ \, \nu}$ and integral~\eqref{BPSEnergy} still make sense. The lack of the natural normalization condition $K_{\mu}K^{\mu}=1$ means that the energy is only defined up to a multiplicative constant, but for any choice of $C$, a minimizer is still required to satisfy Eqs.~\eqref{FOII}. Moreover, it can also be verified that solutions of these equations still solve~\eqref{EL}. These facts and the conservation of~\eqref{BPSEnergy} are sufficient for our purposes.}
Let us now explore some examples of geometries which give rise to the desired behavior of $\xi(r)$ in three or more dimensions.

 Other than Minkowski, conformally flat spacetimes are arguably the simplest class of geometries we can use. Following the notation we used in the $D=1$ case, let $f(r)$ be the conformal factor. From what we have thus far learned about the behavior of $\xi(r)$ in $D\geq 3$, it is clear that the desired behavior for $\xi(r)$ is achieved if $1/f(r)\to C $ as $r\to 0^{+}$ while this same factor must diverge with order $r^{D-2}$ or greater at the boundary. If, in particular, $f(r)\sim r^{D-2}$ asymptotically, then $\xi(r)$ grows logarithmically, while $f(r)\sim r^{n}$ with $n$ an integer greater than $D-2$, gives rise to a power law behavior for this function. One simple class of conformally flat geometries satisfying the above requirements is given by
  \begin{equation}\label{conformalmetric}
 	ds^2=\frac{1}{e^{-1/r}r^n +1}\left[dt^2 - dr^2 - r^2d\Omega^2_{D}\right] \implies \xi(r)=\begin{cases} 
 		\ln(r) + \Gamma\left(-n, \frac{1}{r}\right) +\xi_{0}, & \text{if } D = 2 \\[1ex]
 		\frac{r^{2-D}}{2-D} + \Gamma\left(D-n-2, \frac{1}{r}\right) +\xi_{0}, & \text{if } D \geq 3 
 	\end{cases},
 \end{equation}
where $n\in\mathbb{N}$ and $\Gamma(a,r)$ is the upper incomplete gamma function~\cite{NISTI}.  In the $D=3$ case, this expression can be simplified to $\xi(r)=-1/r + r^{n-1} E_n\left(1/r\right)$, where $E_n$ is the generalized exponential integral. Using the known properties of this function~\cite{NISTI}, we can show that near the origin $r^{n-1}E_n\!\left(1/r\right)
\sim
r^{n}e^{-1/r}\left(1-nr+\cdots\right)$. Hence $\xi(r)\sim -1/r$ $\forall \ n$, as any power of $r$ is suppressed by the exponential. The conformal factor is bounded for all $r$, and changes from unity to zero as the radial coordinate grows from zero to infinity, thus the conformal factor is completely regularized by the exponential and does not give rise to singularities or Killing horizons regardless of $n$. The causal structure is similar to Minkowski spacetime, to which the geometry is identical at small $r$. At infinity the metric degenerates for all $n$, but the physics of spacetime changes significantly with this parameter. In particular, the proper radial distance $R=\int dr(e^{-1/r}r^n +1)^{-1/2}$ grows as  $R\sim r^{1/2}$ and $R\sim \ln(r)$ for $n=1$ and $n=2$ respectively, while for greater values of $n$ this proper length converges to finite values ($R_{\infty}\approx 3.106$ for $n=3$, for example). Thus, the geometry is effectively closed for larger $n$, as a physical observer would measure the spacetime region in which this patch is valid as a finite compact structure. This could be particularly interesting to describe an effective background confining the soliton. The $n=2$ case is a critical point between these two scenarios, as the proper distance is still unbounded, but only grows logarithmically at infinity, with spacelike slices $dt=0$ corresponding to a cylindrical geometry at large $r$.
 
Here, regularization of the field equations again forces $W_{\phi}=0$ at the boundary. At large $r$ one finds, when $n=1$, $\xi(r)\sim \ln r- 0.5772...$, where $ 0.5772...$ is the Euler–Mascheroni constant. For $n=2$ we find $\xi(r)\sim r + \ln(r)$ asymptotically, while for larger $n$ one finds $\xi(r)\sim r^{n-1}/(n-1)+\mathcal{O}\left(r^{n-2}\right)$ in the same regime. \textcolor{black}{Since $\xi\sim -1/r$ and $f(r)\to 1$ as $r\to 0^{+}$, the exponential behavior inherited from the $1D$ model~\cite{bnrt1} as $\xi\to-\infty$ leads to an energy density which vanishes as $e^{-1/r}$ at the origin. At infinity, the energy density decays as $\mathcal{H}\sim e^{-4a\xi(r)}d\xi/dr$. For $n=1$, $\xi$ decays logarithmically, and the energy falls off as $r^{-(4a+1)}$, thus vanishing at the boundary. For $n=2$, $\mathcal{H}$ falls off exponentially as $e^{-4ar}$, while greater powers lead to an even faster suppression as $\mathcal{H}\sim r^{n-2}e^{-4ar^{n-1}}$ }

\begin{figure}[t!]
	\centering
	\includegraphics[width=0.48\linewidth]{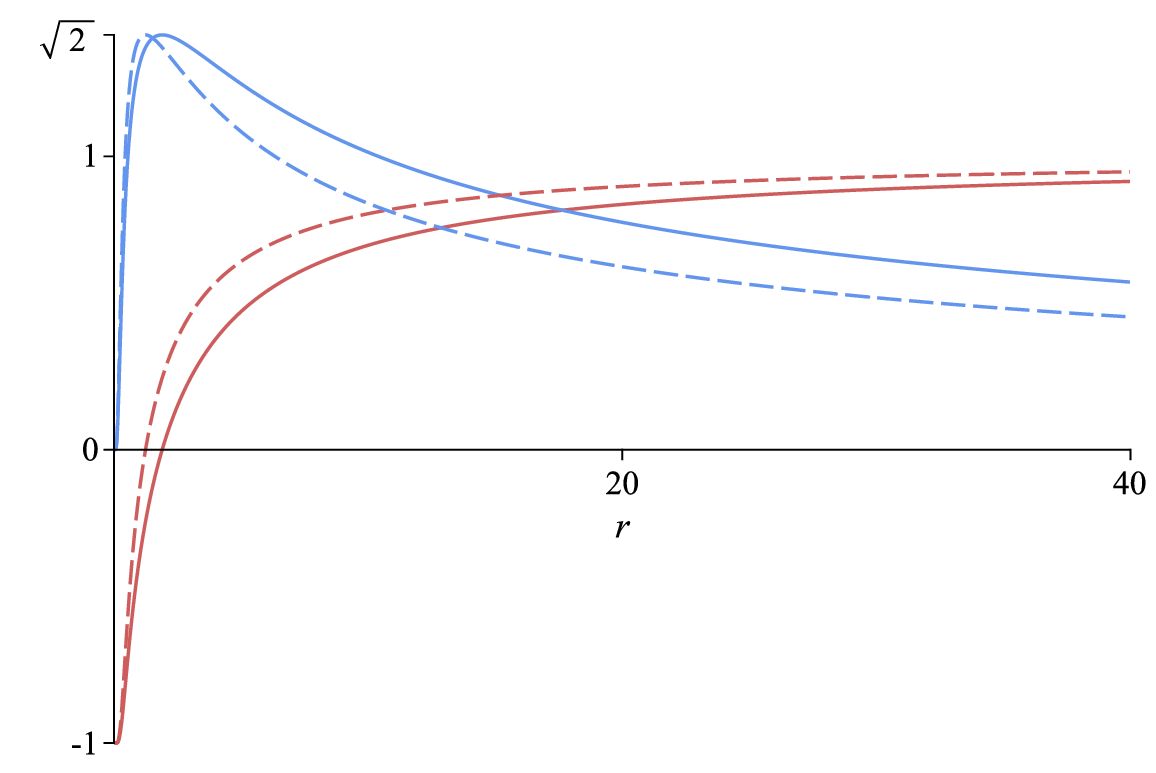}
	\includegraphics[width=0.48\linewidth]{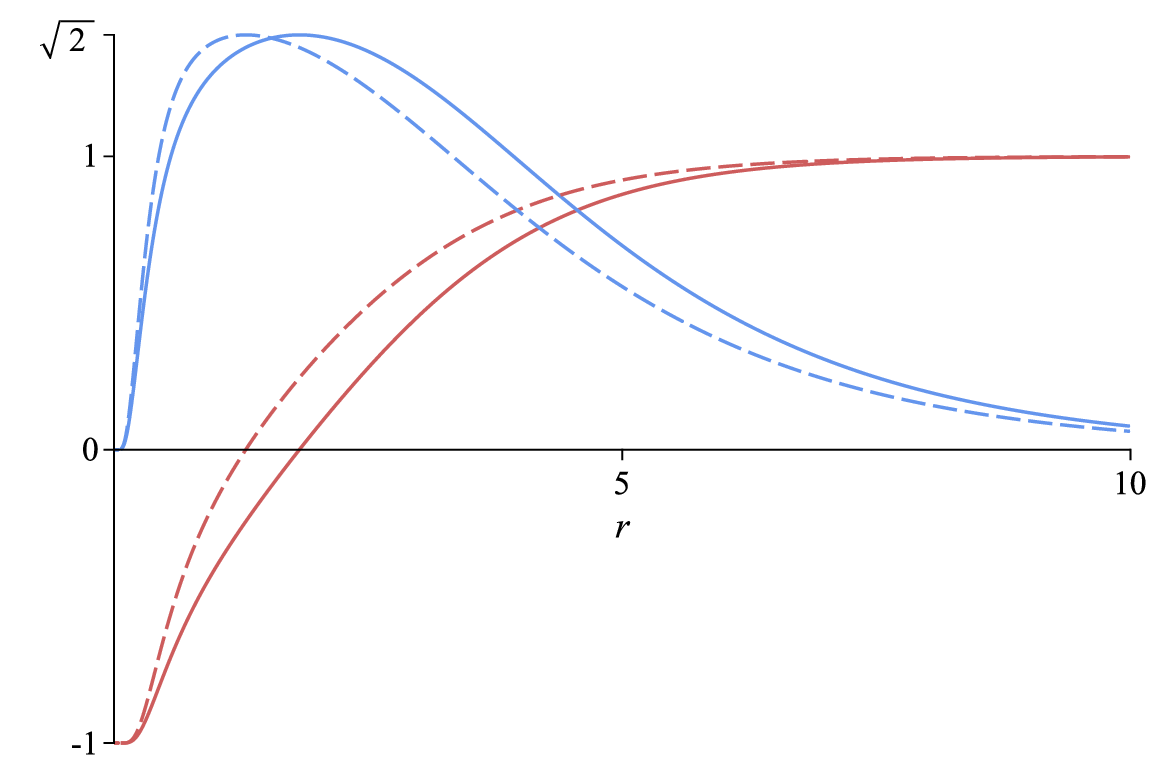}
	\caption{Solutions~\eqref{solutionI} with $a=1/4$  and $\xi(r)=-1/r + r^{n-1} E_n\left(1/r\right)$ for $n=1$ (left) and $n=2$ (right). Solid and dashed lines correspond respectively $\xi_{0}=0$ and  $\xi_{0}=1/2$  in both plots. }
	\label{fig2}
\end{figure}

Evidently, the $r^n$ factor can be easily generalized to an arbitrary function $g(r)$ with a nontrivial power series representation in $r$ at both the origin and infinity, although a closed-form expression for $\xi(r)$ is of course not obtainable for all choices of $g(r)$. In Fig.~\ref{fig2}, we show solutions for this space in $D=3$ for the cases $n=1$ and $n=2$. We see that both solutions reach $\mathcal{M}$ as required, although with largely different asymptotic properties, as the $n=1$ solution displays a significantly larger range. This illustrates the manner in which the tail of the configurations can be changed by the geometry. 

\textcolor{black}{Another spacetime where the aforementioned requirements on $\xi$ are satisfied} is the pure de Sitter (dS) background, which corresponds to the line element
\begin{equation}\label{dS}
ds^2 = (1 - \lambda r^2) dt^2 - \frac{dr^2}{1 - \lambda r^2} - r^2 d\Omega^2 \implies \xi(r) = -\frac{1}{r} + \sqrt{\lambda}\, \text{arctanh}\left(\sqrt{\lambda}\,r\right) +\xi_{0},
\end{equation}
where $\lambda=\Lambda/3>0$ is a scaled cosmological constant and the above metric corresponds to the static patch (see, for example,~\cite{DeSitterSpaceLNs} for a review on the subject). As is well-known, dS spacetime possesses a cosmological horizon corresponding to the surface where the timelike Killing vector $\pb{t}$ becomes null. The energy functional and first-order equations can be consistently defined inside this patch, and we must thus consider the region bounded by $r=0$ and $r_{+}=1/\sqrt{\lambda}$. The factor $B^2(r)/\gamma$ vanishes at both of these values, and regularization thus imposes $W_{\psi}(r\to0^{+})=W_{\psi}(r\to r^{+})=0$ for $\psi=\phi, \chi$ in order to avoid quadratic divergences.
Through~\eqref{dS}, this interval is continuously mapped into $\xi(r)\in(-\infty, \infty)$, thus allowing the fields to reach their vacuum values. Solutions in this geometry can be obtained by direct substitution of~\eqref{dS} into any of the known solutions of the original $D=1$ theory.

\begin{figure}[t!]
	\centering
	\includegraphics[width=0.48\linewidth]{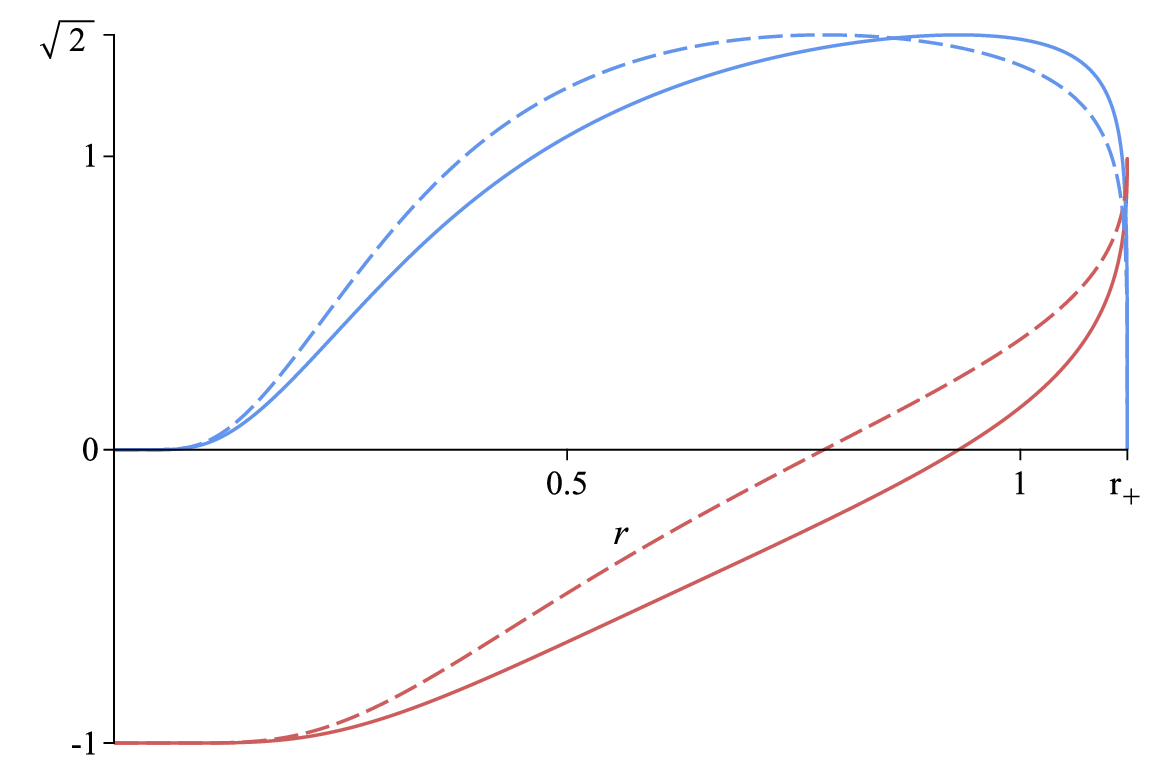}
	\includegraphics[width=0.48\linewidth]{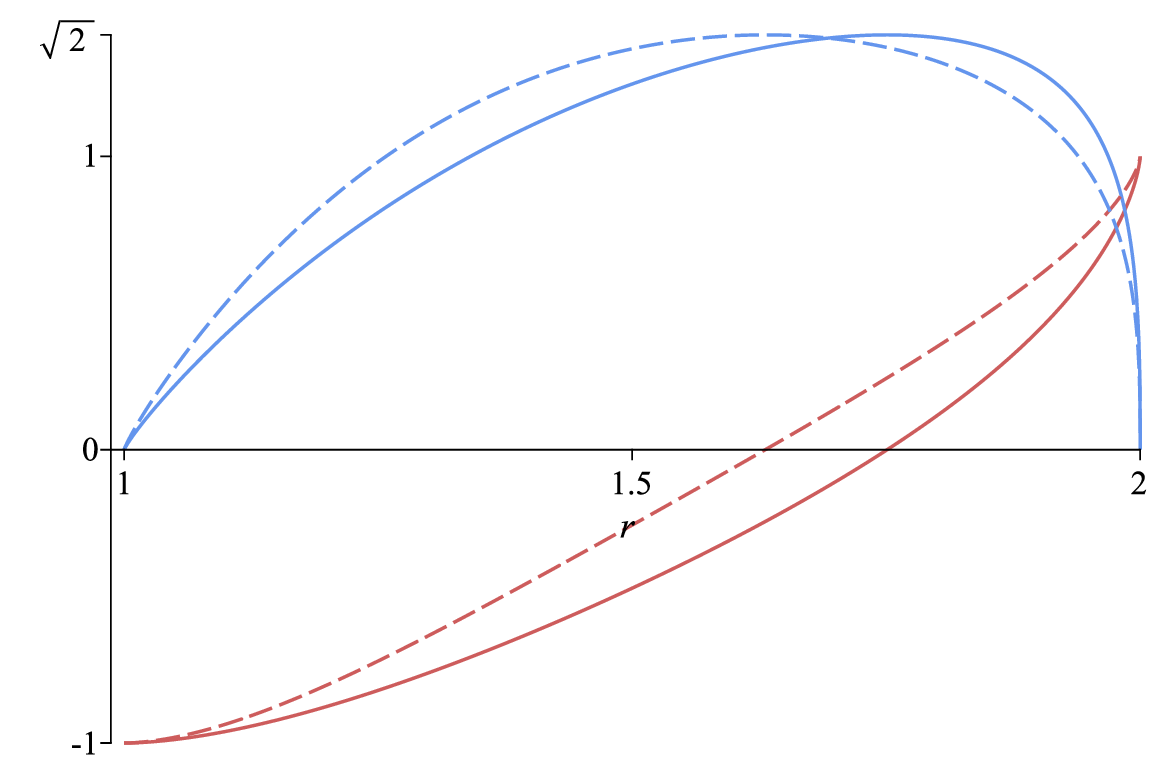}
	\caption{Solutions~\eqref{solutionI} with $a=1/4$  and $\xi(r)$ given by~\eqref{dS} (left) and~\eqref{ScdSxi} (right), which correspond respectively to pure dS and Schwarzschild dS geometries. The metric parameters have been chosen as $\lambda=0.8$  in the left plot and $\mu=3/7, \lambda=1/7$ in the right one. Solid and dashed lines correspond respectively $\xi_{0}=0$ and  $\xi_{0}=1/2$. The cosmological horizons in the left and right plot are located respectively at $r_{+}=1/(2\sqrt{2})\approx 1.118$ and $r_{+}=2$. }
	\label{fig3}
\end{figure}

Both field derivatives fall to zero at the origin, and the energy density vanishes as $e^{-1/r}/r^2$. The situation at the cosmic horizon is, however, very different. The derivatives of $\phi$ and $\chi$ behave respectively as $e^{-\xi}$ and $e^{-\xi/2}$ near the horizon and hence the \textcolor{black}{ energy density behaves asymptotically as} $\sqrt{|g|}g^{rr}\left(\phi'^2(r)+\chi'^2(r)\right)\sim(r_{+}-r)^{\sqrt{\lambda}/2 - 1}$, which can be divergent if $\sqrt{\lambda}/2 < 1$. Nevertheless, this divergence is always integrable since $\sqrt{\lambda}/2 - 1 > -1$ for any choice of $\lambda$, so the divergence does not cause physical inconsistencies. If $ \sqrt{\lambda}/2 \geq 1$ the energy density is everywhere regular, as the power $(r_{+}-r)^{\sqrt{\lambda}/2 - 1}$ becomes a zero (or constant, if equality is exactly attained) instead of a pole. However, a realistic cosmological constant is typically assumed to be very small compared to soliton energy scales, so a mild divergence scenario is arguably more realistic from a physical point of view. \textcolor{black}{Near the origin, no issue arises as the energy density falls to zero as a product between $ d\xi/dr$ and an exponentially decaying factor.} Solution~\eqref{solutionI} with $\xi(r)$ given by~\eqref{dS} is shown in the left-hand side plot of Fig.~\ref{fig3}.

Another viable metric with positive scalar curvature is found in the Schwarzschild de Sitter background, whose line element is written
\begin{equation}\label{SchwarzschilddS}
	ds^2 = \left(1-\frac{2\mu}{r}-\lambda\,r^2\right) dt^2 - \frac{dr^2}{\left(1-\frac{2\mu}{r}-\lambda\,r^2\right)} - r^2 d\Omega^2,
\end{equation}
where $\mu$ is a mass parameter. This metric has both an event horizon and a cosmological horizon, respectively corresponding to the lower and upper roots of $g^{rr}=0$. In order to illustrate the results with simple numerical values for the horizons, let us choose $\mu=3/7,\ \lambda=1/7$, hence $r_{-}=1$ and $r_{+}=2$. Thus
\begin{equation}\label{ScdSxi}
	\frac{d\xi}{dr}= -\frac{7}{r(r-1)(r-2)(r+3)} \implies \xi(r)= -\frac{7}{6}\ln r
	+\frac{7}{4}\ln|r-1|
	-\frac{7}{10}\ln|r-2|
	+\frac{7}{60}\ln|r+3|
	+\xi_{0}.
\end{equation}

Solutions thus approach the vacua with a power law behavior as the horizons are approached. The field derivatives vanish at the inner boundary (i.e., the event horizon) and diverge at the cosmological horizon. Solution~\eqref{solutionI} with $\xi(r)$ as above is shown as the right-hand plot of Fig.~\ref{fig3}. For the chosen values of the parameters, the energy density grows as a power of $(2-r)^{-3/10}$ at this boundary, which again corresponds to an integrable singularity. As in pure de Sitter spacetime, we could eliminate this pole with convenient choices of the constants. For general choices of the parameters, this happens when $(3\lambda r_{+}^2-1)r_{+}\leq 1$. Since $r_{+}$ also depends on $\mu$, this inequality can be satisfied even for small values of $\lambda$, although this would require some fine-tuning of the black hole mass. \textcolor{black}{The  near-horizon behavior is largely dictated by the quantities $\kappa_{\pm}$ obtained through the power expansion $1-2\mu/r-\lambda\,r^2\approx \kappa_{\pm}\left|r-r_{\pm} \right|$. Indeed, it may be verified that $\frac{d\xi}{dr}\big|_{r_{\pm}}\approx 1/\left(\kappa_{\pm}r_{\pm}^2\right)$. The physical significance of $\kappa_{\pm}$ lies in the fact that they are proportional to the surface gravity of each horizon, thus giving a physical interpretation to the map which defines $d\xi/dr$, and explaining the aforementioned role played by $\xi(r)$ on the energy density near each horizon.}

\subsection{Models with non canonical derivative term $(P(\phi,\chi),Q(\phi,\chi)\neq 1)$}\label{subsec:noncanonical}

We may now move on to the generalized theories corresponding to nontrivial choices of $P(\phi,\chi)$ or $Q(\phi,\chi)$. The simplest such class of models is obtained when these functions depend on only one of the fields, say $\chi$, and $W$ can be written in the form $W(\phi,\chi)=w_{1}(\phi)+w_{2}(\chi)$. In this case the first-order equations are \emph{separable} in the sense that the second equation in~\eqref{FOxi} can be integrated independently and the result substituted into the $\phi$ equation as a function $\chi(r)$, thus allowing the equations to be solved by quadrature. Moreover, the potential can only vanish if $\left(w_{1}\right)_{\phi}=\left(w_{2}\right)_{\chi}=0$, so, \textcolor{black}{for nonsingular choices of $P$ and $Q$,} the vacuum manifold of the theory can be decomposed in the form $\mathcal{M}=\mathcal{M}_{\phi}\times\mathcal{M}_{\chi}$. \textcolor{black}{When $P$ or $Q$ has zeroes, this statement must be understood in the generalized sense discussed above: points where the potential or the field equations are undefined cannot be identified with true vacua, but may belong to an extended set of admissible limiting boundary values if they are consistent weak solutions with finite energy. If an $\epsilon$-regularization scheme is possible, the elements of the generalized set $\mathcal{M}$ in singular theory can also be seen as limits of true vacua of an $\epsilon$-regularized family, since the potential is well-defined for any $\epsilon\neq 0$ when such a family exists.} Such models have been investigated in flat two-dimensional spacetimes in Ref.~\cite{liao1} and follow-up works. Let us assume, for definiteness, that we are dealing with a separable model such that $Q=1$ and $P=P(\chi)$ and let us consider the upper sign equations in~\eqref{BPSII} for the remainder of this work. Then $\chi$ is easily obtained through direct integration of $d\chi(\xi)/d\xi=W_{\chi}(\chi)$. The remaining equation can thus be written
\begin{equation}\label{xitilde}
	\deriv{\phi(\tilde\xi)}{\tilde\xi}=W_{\phi}(\tilde\xi), \quad \quad \text{where} \quad \quad  \tilde\xi\equiv \int  \frac{d\xi}{P(\chi(\xi))} + \tilde{\xi}_{0}.
\end{equation}
The $\tilde\xi$ factor may alternatively be written in the form $\tilde\xi=\int dr\bigl[ B^2(r)/\left(\gamma(r) P(\chi(r))\right)\bigr],$ so the scalar coupling leads to an effective geometric factor $\tilde\xi$. The solution behaves as a kink-like (or domain wall) single-field defect in an effective geometry with line element $ds^2=A^2(r)dt^2 -\bigl[\tilde B^2(r)(dr)^2 + \rho^2(r)d\Omega^2_{D}\bigr]$, where $\tilde B(r)\equiv B^2(r)P^{-2}(\chi(r))$. The scalar coupling hence acts as a pointwise transformation in $g_{rr}$. The properties of $\phi(\tilde\xi)$ thus emerge from an interplay between the physical background geometry and the \qt{geometric} constriction enacted by the $\chi$ field, which maps $\xi$ into $\tilde\xi$. 

\begin{figure}[t!]
	\centering
	\includegraphics[width=0.48\linewidth]{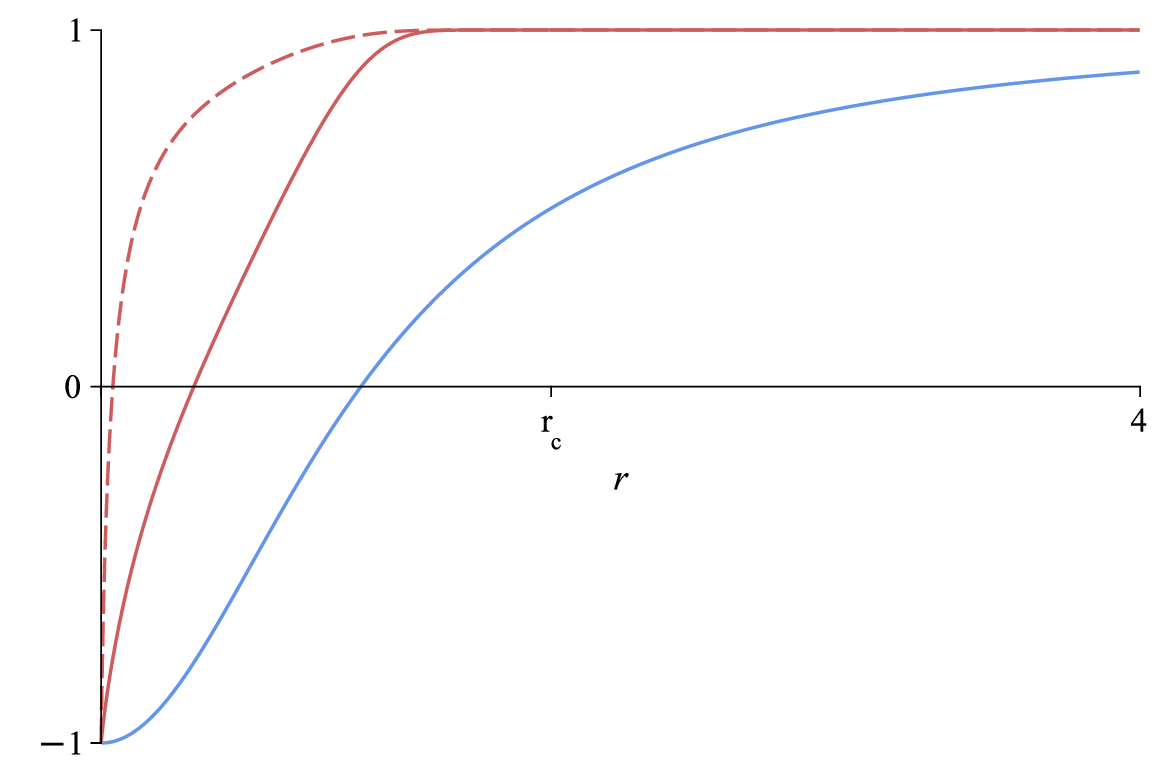}
	\includegraphics[width=0.48\linewidth]{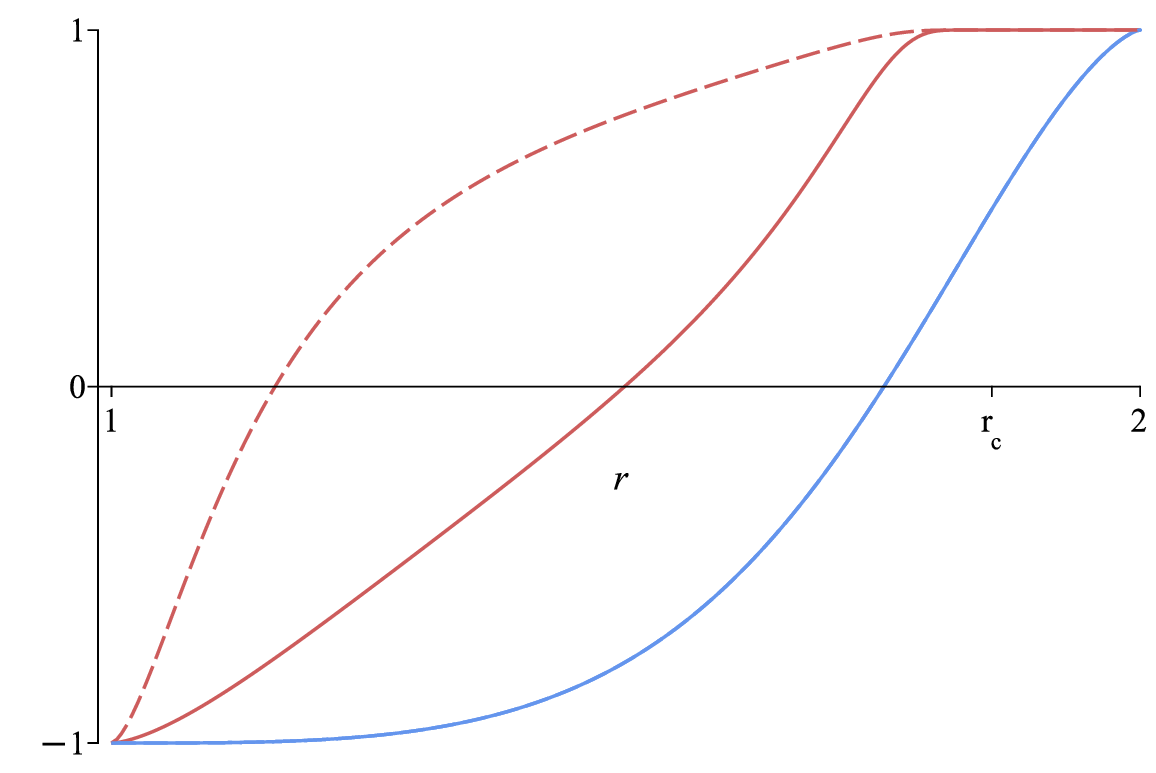}
	\caption{Solutions $\phi(r)=\tanh(\tilde\xi(r))$ for $r\leq r_{c}$, $\phi=1$ for $r\geq r_{c}$ (red), $\chi(r)=\tanh(\xi(r))$ (blue) of~\eqref{BPSII} with $W(\phi,\chi)$ and $P$ given by~\eqref{SPseparable} for $p=1$, $\epsilon=-1/2$. The left plot corresponds to flat $D=2$ space with $\xi(r)=\ln(r)$, while the right-hand plot corresponds to Schwarzschild dS spacetime~\eqref{SchwarzschilddS} with $\xi(r)$ given by~\eqref{ScdSxi}. Solid and dashed lines correspond respectively $\tilde\xi_{0}=0$ and  $\tilde\xi_{0}=1$, while $\xi_{0}=0$ in both plots.}
	\label{fig4}
\end{figure}

We may use the generalization engendered by $P(\chi)$ to modify the model in order to give rise to topological solutions that would otherwise not be possible in asymptotically flat spacetimes. To this end, let us consider the family of models generated by functions:

\begin{equation}\label{SPseparable}
W(\phi,\chi) = \phi -\frac13\phi^3 +  \chi\frac{|\chi|^{1/p+1}}{2+1/p} -  \chi\frac{|\chi|^{-1/p+1}}{2-1/p}, \quad \quad P_{\epsilon}(\chi)=(\chi+\epsilon)^2.
\end{equation}
where $p$ is a parameter which we take as an odd positive integer, $\epsilon\in(-1,1)$, and we have written $|\chi|$ in $W(\phi,\chi)$ in order to unambiguously specify the real branch of these fractional powers. This function is a sum of the superpotential of a $\phi^4$ theory, to which the Lagrangian is reduced when $\chi$ is fixed at a vacuum value, and that of the $p$ model introduced in Ref.~\cite{PRL2003}. \textcolor{black}{If $\epsilon\neq0$,} these choices give rise to a potential that vanishes if $\phi=\pm 1$ and $\chi=0,\pm 1$, except for $p=1$, where the minima $(\pm 1,0)$ do not exist. \textcolor{black}{In the limiting case $\epsilon=0$, the points $(\pm 1,0)$ are not true vacua, and must instead be interpreted as limiting boundary values in the weak sense discussed above, or as limits of true vacua in the sense of an $\epsilon$-regularization obtained by treating $\left.P\right|_{\epsilon=0}$ as the $\epsilon\to 0$ limit of the $P_{\epsilon}$ family defined as in~\eqref{SPseparable} }. Note that the $Z_{2}$ symmetry is in general broken outside of $\chi=0$ orbits due to the introduction of $P(\chi)$. This symmetry is however recovered in the limit $\epsilon\to0$. The independent $d\chi(\xi)/d\xi=W_{\chi}(\chi)$ equation is solved, in an appropriate neighbourhood, by 
\begin{equation}\label{pmodelSol}
\chi(\xi)=\tanh^p\left(\frac{\xi}{p}\right).
\end{equation}
The validity and extension of the above $\chi$ configurations are dependent on the choices of $\xi_{0}$, $p$, and the background geometry. If $p=1$, $\chi$ is the standard $\tanh(\xi)$ solution of the $\chi^4$ model, while $\phi=\tanh(\tilde\xi)$, where
\begin{equation}
\tilde{\xi}=\frac{\ln\left(\left|\tanh\left(\xi\right) + 1\right|\right)}{2\left(\epsilon^2- 4{\epsilon} + 2\right)} 	-\frac{2{\epsilon} \ln\left(\left|\tanh\left(\xi\right) + {\epsilon}\right|\right)}{{(\epsilon}^{2}-1)^2} - \frac{\ln\left(\left|\tanh\left(\xi\right) - 1\right|\right)}{2{\epsilon}^{2} + 4{\epsilon} + 2} + \frac{1}{\left({\epsilon}^{2} - 1\right) \left(\tanh\left(\xi\right) + {\epsilon}\right)} - \tilde\xi_{0},
\end{equation} 
This solution can reach the \textcolor{black}{ boundary values necessary for topological solitons} in any background where $\xi$ ranges from $-\infty$ to $\infty$. Since $\chi$ interpolates between $-1$ and $1$, there must exist a point $r_{c}$ such that $\chi(r_c)=\epsilon$ and thus $P(\chi)=0$.  At $r=r_c$ the term $1/P(\chi)$ introduces a pole of order $2$ into~\eqref{BPSphi}, which must be compensated by a zero of equal or greater order at $W_{\phi}$ to ensure regularity. Solutions in the $(-1,-1)\to (1,1)$ sector can be divided into two classes, the first of which is $\phi(r)=\tanh(\tilde\xi(r))$ for $r\leq r_{c}$, $\phi=1$ for $r\geq r_{c}$. This construction ensures continuity of both the solution and its derivative. Thus, $\phi$ reaches its boundary value at a finite distance from the origin even if $\xi$ is unbounded from above and below. In this case, the $\phi$ profile is similar to that of compacton~\cite{Compactons,CompactonsII, CompactonsIII} solutions, in the sense that its derivative has compact support. The remaining solutions are such that $\phi=-1$ for $r\leq r_{c}$, $\phi(\tilde\xi(r))=\tanh(\tilde\xi(r))$ for $r\geq r_{c}$, which are also unusual due to $\phi$ being constant inside a closed set of finite volume. The full solution is however not strictly a compacton in either case, since the $\chi$ profile can only reach its boundary value at infinity if $p=1$. \textcolor{black}{The energy density at the domain endpoints is given by $\mathcal{H}\sim \left(e^{\pm 2\xi}+e^{\pm 2\tilde\xi}\right)d\xi/dr$. The flat $D=2$ solution vanishes as $r^{-3}$ as $r\to\infty$, while the near-origin behavior is dominated by the $\phi$ field if this field is compact, and gives rise to an energy density which vanishes as the fractional power of $r^{7/9}$. On the other hand, the Schwarzschild de Sitter solution depicted in the right-hand side of Fig.~\ref{fig4} gives rise to a behavior of the form $\mathcal{H}\sim (r-r_{+})^{9/5}$ near the cosmological horizon located at $r_{+}=2$, which comes entirely from the $\chi$ field since $\phi'$ has compact support. Near the event horizon, given by $r_{-}=1$, one finds $\mathcal{H}\sim (r-r_{-})^{6}$. Thus, the energy density reaches zero smoothly in this space.   }

The solutions discussed in the previous paragraph are valid for any of the backgrounds used in previous examples, since $\xi$ is unbounded for all of these geometries.  In Fig.~\ref{fig4}, we depict the \qt{compacton profile} solutions for $\epsilon=-1/2$ in flat $D=2$ and Schwarzschild dS ($\lambda=\mu/3=1/7$) spacetimes. If $\xi_{0}=0$, the derivative support ends at $r_{c}=\sqrt{3}$ and $r_{c}\approx 1.856$ respectively for these backgrounds. The constriction induced by $P(\chi)$ into the $\phi$ field  has a very significant effect on the size of the soliton, as $\phi(r)$ approaches its boundary value very quickly despite the logarithmic character of $\xi$ in both backgrounds. We also note a significant change in internal structure of the domain wall when the zero-mode parameter $\tilde\xi_{0}$ is changed, as illustrated by the difference between solid and dashed lines in the figure. Similar changes can be achieved through variation of $\xi_{0}$, which represents the remaining zero mode parameter.  

As in the BNRT model, it can be shown that $\chi$ cannot interpolate between $\pm 1$ if $\xi$ is bounded from above or below, thus precluding topological solutions in asymptotically flat spacetimes with $D\geq 3$ when $p=1$. However, the potential  gains additional  zeroes $(\pm 1, 0)$ for greater values of $p$. These new minima are fundamentally different from the others because $\chi(\xi_{0})=0$ can be satisfied nontrivially for finite $\xi_{0}$, thus allowing for a solution compatible with~\eqref{EffectiveBCS} in asymptotically flat spacetimes of any dimension. In $D=1$, Eq.~\eqref{pmodelSol} can be extended to all $x$, giving rise to a configuration which interpolates between three minima. This is the one-field solution originally given in~\cite{PRL2003}, where it was classified as two-kink configuration. However, single kink configurations also exist in the model, and correspond to semi-compact profiles which equal $\tanh^p\left((x-x_{0})/p\right)$ within either of the intervals $(-\infty,x_{0}]$ or $[x_{0}, \infty)$, and are constant elsewhere. It is straightforward to extend all of these results to any background geometry where $\xi$ ranges from $-\infty$ to $\infty$. 

In flat $D=3$ spacetime, $\chi$ can only interpolate between two vacuum values. For definiteness, let us search for solutions connecting $\chi=-1$ to $\chi=0$. The field approaches $\tanh^p(\xi_{0}/p)$ asymptotically, so the boundary conditions can be satisfied provided $\xi_{0}\in[0,\infty)$. The $\chi$ profile is compact in general, vanishing outside a ball of finite radius. The only exception to this rule is the $\xi_{0}=0$ configuration, which is non-compact since $\xi$ only vanishes at infinity. For a solution satisfying~\eqref{EffectiveBCS} in an arbitrary asymptotically flat space, there exists, for every solution with $\xi_{0}\neq 0$ , a radius $r_{c}$ such that $r>r_{c}\implies W_{\chi}=0$. In flat $D=3$ and Schwarzschild backgrounds, $r_c$ and $\xi_{0}$ are related by
\begin{equation}\label{rc}
	\xi_{0}=\begin{cases}
		&1/r_{c}\quad\quad\quad\quad\quad \text{ ($D=3$ Minkowski spacetime)},\\
		&\ln\left(\frac{r_c}{r_c-1}\right) \quad\quad \quad \text{(exterior Schwarzschild background)}.
	\end{cases}
\end{equation}

\begin{figure}[t!]
 	\centering
 	\includegraphics[width=0.48\linewidth]{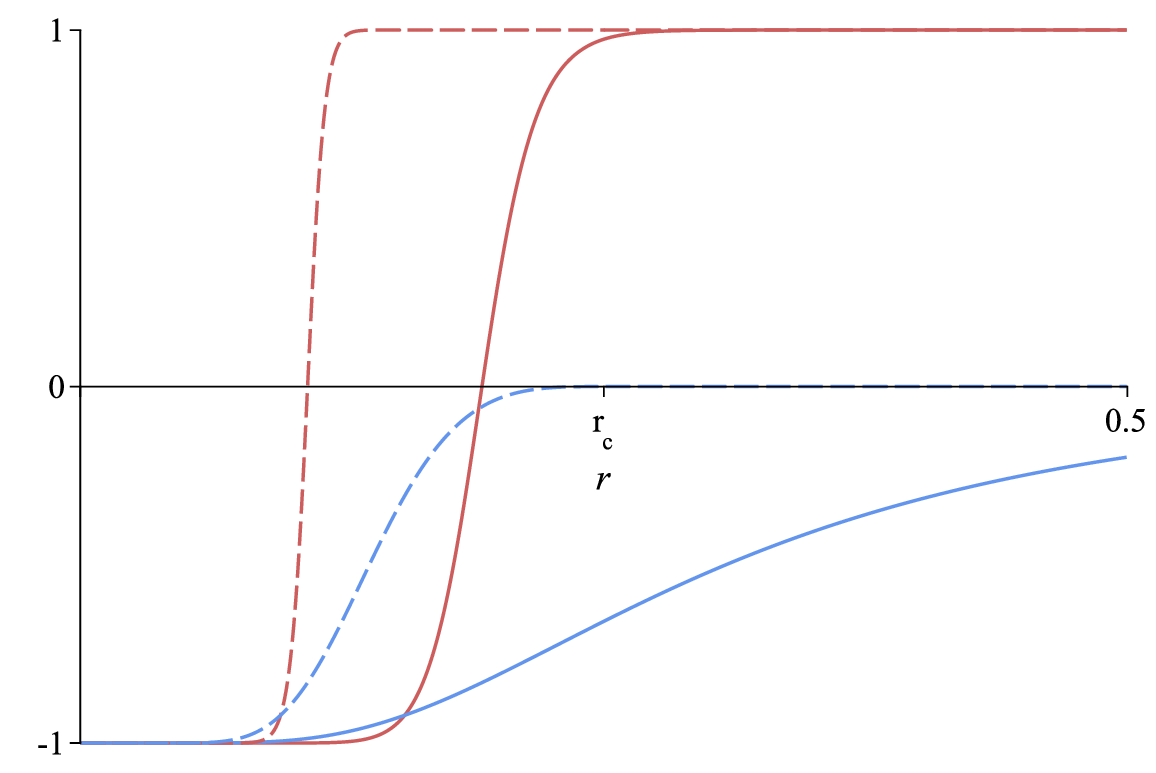}
 	\includegraphics[width=0.48\linewidth]{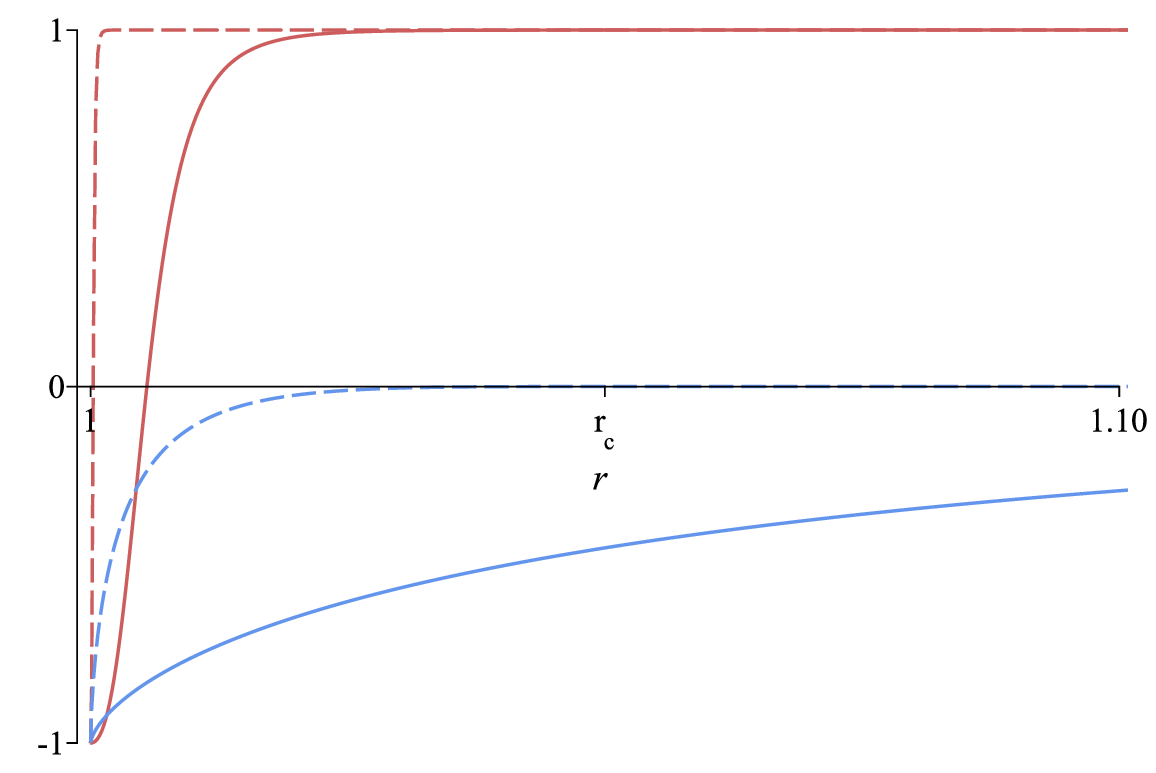}
 	\caption{Solutions $\phi(r)$ (red), $\chi(r)$ (blue) of~\eqref{FOII} for $W(\phi,\chi)$ and $P$ given by~\eqref{SPseparable} with $p=3$, $\epsilon=0$. The left-hand plot corresponds to flat $D=3$ space, while the right-hand plot corresponds to an exterior Schwarzschild spacetime~\eqref{Sch} with unit Schwarzschild radius. Solid lines correspond to \qt{noncompact} profiles $r_{c}=\infty$ while dashed lines represent solutions that reach their boundary values at $r_{c}=1/4$ (left) and $r_{c}=1.05$ (right) while $\tilde\xi_{0}=0$ in both plots.}
 	\label{fig5}
 \end{figure}

We now turn our attention to the remaining scalar field. It should be clear from our previous results that we must look for configurations which display a compact or half-compact profile when viewed as functions of $\xi$ if $\phi$ is to be consistent with~\eqref{EffectiveBCS}. To this end, $P(\chi)$ must vanish at some point of the domain, which, in the sector $(-1,-1)\to (1,0)$ we are currently considering, is possible if $\epsilon\in (-1,0]$. Similarly, a solution in the sector $(-1,0)\to (1,1)$ requires $\epsilon\in [0,1)$. This distinction reflects the breaking of $Z_{2}$ symmetry in the model, which is why $\epsilon=0$ is the only scenario where the equivalence between these sectors is restored. Solutions can be found explicitly through direct integration of~\eqref{xitilde}, which can be written for any $\epsilon$, although its closed-form expression for an arbitrary choice of this parameter is prohibitively large, and shall thus be omitted. The expression is greatly simplified in the limiting case $\epsilon\to 0$, where $\phi$ is written
 \begin{equation}\label{compactphi}
\phi(\tilde\xi)=\tanh(\tilde\xi), \quad \quad \text{for} \quad \quad \tilde\xi=\xi-3\coth\left(\frac{\xi}{3}\right)\left[1+\frac{\coth^4\left(\frac{\xi}{3}\right)}{5}+\frac{\coth^2\left(\frac{\xi}{3}\right)}{3}\right] +\tilde{\xi_{0}}.
 \end{equation}
The full solution is thus specified by the two constants $\xi_{0}\in[0,\infty)$ and $\tilde{\xi_{0}}\in (-\infty,\infty)$, which together parameterize its two-dimensional moduli space. We thus have solutions interpolating between the values $(-1,-1)$ and $(1,0)$ in $D=3$ Minkowski and Schwarzschild backgrounds. If $\xi_{0}\neq 0$, the $\phi$ configuration is given by~\eqref{compactphi} inside a compact set of finite volume, and is unity otherwise. Thus, both fields, and hence the solution itself, present a compacton profile. The energy density vanishes for $r>r_{c}$ so the defect is completely localized within the region bounded by this radius. The remaining constant $\tilde{\xi_{0}}$ changes the profile of the solution as a zero mode, but does not induce compactification. In Fig~\ref{fig5}, we depict these solutions for two values of $\xi_{0}$, both in flat $D=3$ and Schwarzschild backgrounds. \textcolor{black}{In flat $D\geq 3$ spacetime, the energy density always vanishes at the origin because of the exponential falloff which emerges when the arguments of $\tanh(\xi)$ and $\tanh(\tilde\xi)$ go to negative infinity. For the near-horizon solution in Schwarzschild spacetime, one finds $d\xi/dr\sim 1/(r-r_s)$, so the energy density $\mathcal{H}\sim \left(e^{4\xi/p}\right)d\xi/dr\sim (r-r_s)^{4/p-1}$, which is always integrable since the exponent is dominated by $(r-r_s)^{4/p-1}$ for all $p$. At infinity the energy density vanishes even faster than in the analogous $D=1$ model, since $d\xi/dr$ also vanishes in asymptotically flat spaces of higher dimensions.}
 
We note that, when $\epsilon=0$, the potential is undefined at $(\phi,\chi)=(\pm 1, 0)$, so strictly speaking these values cannot be identified with vacua. As remarked above, solutions can also be found for other values of $\epsilon$, so the theory can be seen as the $\epsilon\to 0$ limit of a family of Lagrangians specified by $P_{\epsilon}$, $\epsilon\in(-1,0)\cup(0,1)$, each of which possesses true vacua at $(\phi,\chi)=(\pm 1, 0)$. Moreover, the potential approaches a zero along any trajectory which solves the appropriate weak formulation of the Euler-Lagrange equations. It is thus straightforward to extend the definition of $\mathcal{M}$ to accommodate the $\epsilon=0$ case, \textcolor{black}{ which can be achieved by defining this manifold as an extended set of admissible limiting boundary values inherited from the true vacua of the $\epsilon\neq 0$ regularized family.}

The functions $P$ and $Q$ of the generalized models may also be used to modify theories in which the superpotential is not additive. Unlike the separable framework, the introduction of auxiliary functions does not lead to equations that can be generally solved by quadrature, and an integrating factor is usually required to solve the first-order equations. Nevertheless, these functions may still cause significant changes to the physical system, such as giving rise to new minima or allowing for regular solutions in certain BPS sectors. As we have seen, the BNRT model with usual derivative terms does not support topologically nontrivial solutions in asymptotically flat spacetimes of four or more dimensions. One could theoretically choose the generalized functions in such a way as to avoid this problem, since $P$ and $Q$ appear in the potential and can create new minima or generate compact structures. This can be achieved if these functions depend nontrivially on both fields (although, in this case, the boundary conditions are not determined exclusively by the superpotential, and must be changed), but it may be simpler to satisfy these requirements with the introduction of a new model obtained by suitable choices of both $W(\phi,\chi)$ and the generalized derivative terms. To this end, consider a potential of the form

\begin{equation}
	V(\phi,\chi,r)=\frac{B^2(r)}{2\gamma(r)}\left[ |\phi|^{\frac{1}{p}}\left(|\phi|^{\frac{2p-3}{2p}}-|\phi|^{\frac{2p+1}{2p}}-\frac{(1-2a)\beta^2}{p^2}\chi^2\right)^2  +\frac{2a^2\chi^2}{p^2|\phi|^{\frac{2p-1}{2p}}}\left(|\phi|^{\frac{2p+1}{2p}}-\alpha\beta^2\chi^2\right)^2\right],
\end{equation}
where $a$ is a real parameter and we have defined $\alpha\equiv(2p-3)/(4p^2) $ and $\beta\equiv\sqrt{\frac{a(2p+1)}{2(1-2a)}}$. This potential is derived from a superpotential (shown below) which combines features of the BNRT and $p$ models considered above, together with the choices $P=1$, $Q(\phi)=|\phi|^{(2p-1)/(2p)}$. As in the previous example, $(0,0)$ does not correspond to a true vacuum due to the vanishing of $Q(\phi)$ at this point, but the same reasoning from before can be used to show that solutions exist in the weak sense or as a limit of a family of solutions derived from $Q_{\epsilon}(\phi)=|\phi|^{{(2p-1)/(2p)}}+\epsilon^2$. The potential vanishes at $(\phi,\chi)=(\pm 1,0)$ and,  \textcolor{black}{if $\epsilon=0$ exactly, the potential approaches zero (being thus a limiting boundary value consistent with the finite energy assumptions)} for any orbit such that the limit $(\phi,\chi)\to (0,0)$ exists. Moreover, when $a\in(0,1/2)$, there are four additional minima made up from the combinations of $\phi=\pm\big(\frac{2p-3}{2p+1-8a}\bigr)^{p/2}$ and $\chi=\pm\frac{1}{\beta\sqrt{\alpha}}\bigl(\frac{2 p-3}{2 p+1-8a}\big)^{(2 p+1)/8}$. The potential is singular along most of the line $\phi=0$, thus forcing all physical orbits that cross this line to pass through the origin.  

Due to the use of the absolute value function, the potential cannot be derived from a globally defined auxiliary function. However, a superpotential construction is possible with use of two patches corresponding to each hemisphere, namely,
\begin{equation}
	W^{\sigma}(\phi,\chi)=\sigma\left\{\cfrac{p}{2p-1}\,(\sigma\phi)^{\frac{2p-1}{p}}-\cfrac{ p}{2p+1}\,(\sigma\phi)^{\frac{2p+1}{p}}+\cfrac{a(\sigma\phi)^{\frac{2p+1}{2p}}}{p}\chi^2+\cfrac{a\beta^2(2p-3)\chi^4}{8p^3}\right\},
\end{equation}
where $\sigma=\text{sign}(\phi)$. The two branches connect at the origin and thus give rise to consistent BPS equations throughout the acceptable portion of field space. Let us first work in the positive branch and try to find orbits valid for $\phi>0$. The one-form in Eq.~\eqref{orbiteq} is not exact for this superpotential, but we may attempt to find an integrating factor by using the ansatz	$\mathcal{I}=\phi^m\chi^n$, where $m$ and $n$ are real constants to be determined through the requirement of consistency with the exactness condition. Proceeding in this manner we find that the integrating factor $\mathcal{I}=\phi^{(1-p)/p}\chi^{-\left(a+1\right)/a}$ suffices to integrate~\eqref{orbiteq} and obtain the general orbit:
	\begin{equation}\label{orbitnewmodel}
		(\sigma\phi)^{\frac{p-1}{p}}-(\sigma\phi)^{\frac{p+1}{p}}-\frac{\beta^2(\sigma\phi)^{\frac{1}{2p}}\chi^2}{p^2}+C(\sigma\phi)^{\frac{p-1}{p}}|\chi|^{\frac{1}{a}}=0.
	\end{equation}
\cmmnt{	\begin{equation}
		\left(\frac{3}{2}+p\right)=0 \implies p=-\frac{3}{2} ; \quad -3\left(\frac{13}{6}+p\right)=2a\left(q+1\right) \implies q=-\left(\frac{a+1}{a}\right),
	\end{equation}}
Outside of the line defined by $\chi=0$, the above orbit can be written in the more compact form $|\chi|^{-1/a}\left\{p^2\left(1-(\phi^2)^{1/p}\right)-\beta^2(\sigma\phi)^{3/(2p)-1}\chi^2\right\}=C_{2}$.
Choosing $C=0$, which is in particular consistent with the sector $(-1,0)\to (0, 0)$, we can substitute this orbit into the first-order equations to derive the solution
\begin{equation}\label{finalSol}
	\phi(\xi)=\tanh^p\left(\frac{2a\xi}{p}\right), \quad \quad \quad \quad  \chi(\xi)=\frac{p}{\beta}\tanh\left(\frac{2a\xi}{p}\right)\left|\tanh\left(\frac{2a\xi}{p}\right)\right|^{\frac{2p-7}{4}}\sech\left(\frac{2a\xi}{p}\right).
\end{equation}

\begin{figure}[t!]
	\centering
	\includegraphics[width=0.48\linewidth]{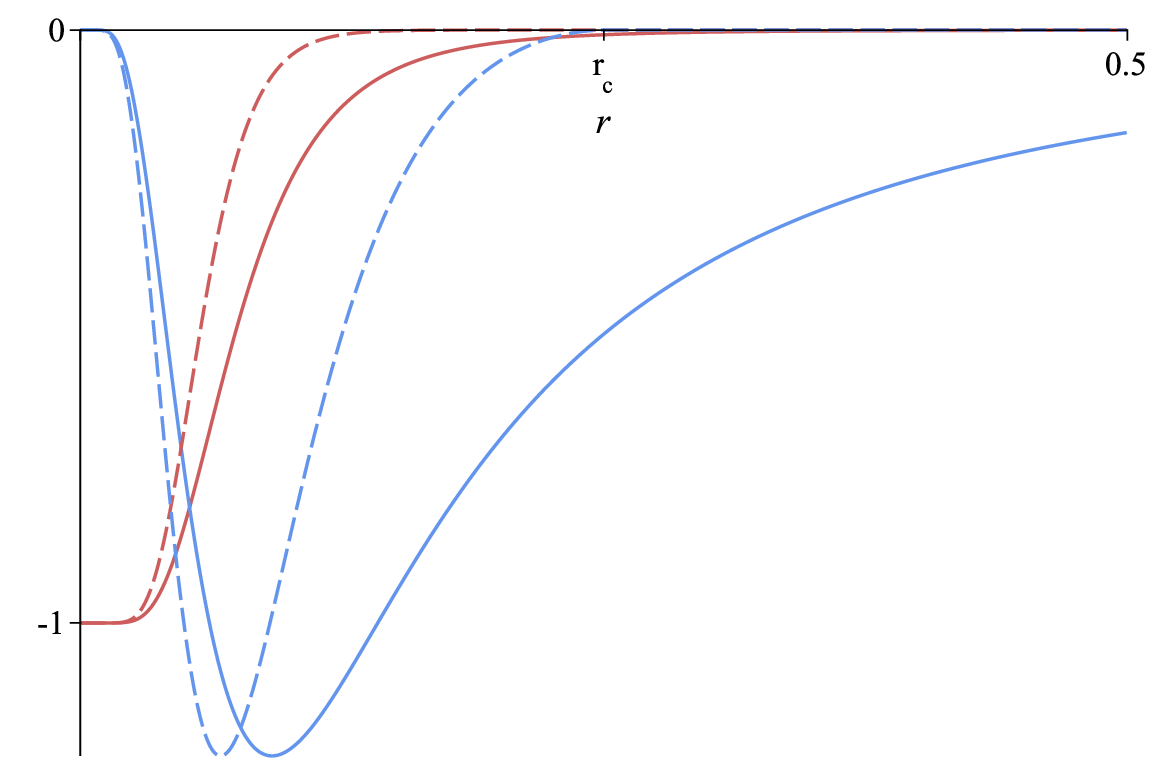}
	\includegraphics[width=0.48\linewidth]{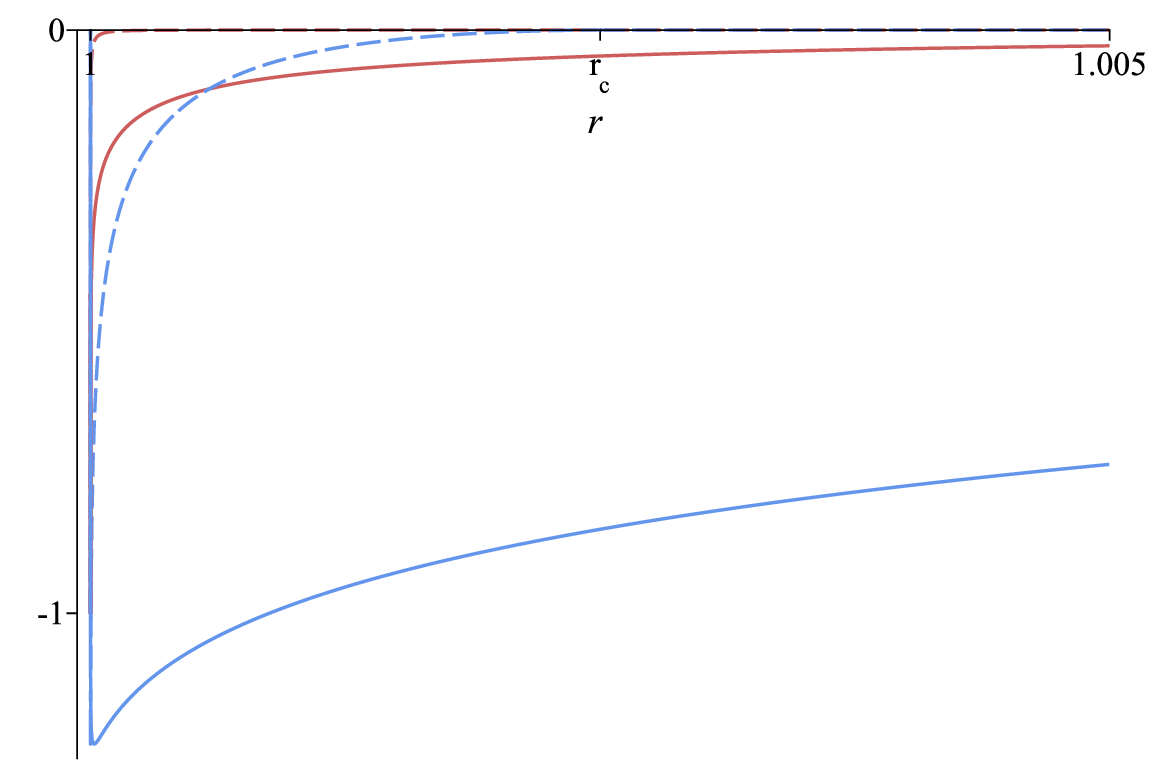}
	\caption{Solutions $\phi(r)$ (red), $\chi(r)$ (blue) given by~\eqref{finalSol} for $p=5$ and $a=1/4$. Solutions are shown in flat $D=3$ (left) and Schwarzschild (right) geometries, with the Schwarzschild radius taken as unity. Solid lines correspond to $\xi_{0}=0$ in both plots, while dashed lines represent profiles that are compactified at $r_{c}$ according to~\eqref{rc}. The compactification radius is $r_c=1/4$ in the left plot and  $r_c=1.0025$ in the right one.}
	\label{fig6}
\end{figure}

This solution is shown in Fig.~\ref{fig6} in the case $p=5$, both in flat $D=3$ and Schwarzschild backgrounds.  As was the case in the previous example, solutions with $\xi_{0}>0$ compactify at a finite radius $r_{c}$, which is found as the positive root of the equation $\phi(r_{c})=0$. The relationship between $\xi_{0}$ and $r_{c}$ is still given by~\eqref{rc} in these spacetimes, regardless of the choice of $p$. Because of the higher power $p=5$ used in these plots, solutions have an overall shorter range when compared to our previous examples. In particular, the compact solution in the Schwarzschild background has an energy density with support from $r=1$ to $r=r_c=1.0025$, which amounts to only $0.25\%$ of the Schwarzschild radius in the example. The size of the soliton decreases monotonically with $\xi_{0}$, so this parameter can be freely changed to shrink or enlarge the defect. The choice $\xi_{0}\approx 0.693$, for example, implies $r_{c}\approx 2$ and hence leads to a defect roughly of the size of the black hole. Arbitrarily small, positive choices of this parameter give rise to energy densities with arbitrarily large support, while greater values can be used to confine the soliton inside a progressively smaller neighbourhood of the event horizon. In practice, even the $\xi_{0}=0$ defect differs appreciably from \textcolor{black}{the limiting boundary configuration} only within a finite region determined by the power-law character of its tail, but the $\xi_{0}>0$ examples are fundamentally different in that the energy density is exactly zero. \textcolor{black}{Since $d\xi/dr\to 0$ as $r\to\infty$ in asymptotically flat spaces, while $\xi$ is bounded in the same regime, the energy density vanishes asymptotically (trivially so in the noncompact limit $\xi_{0}=0$). Near the origin, the energy densities behave as $\mathcal{H}\sim e^{-\frac{4a\xi}{p}} d\xi/dr$. In flat space, $\mathcal{H}\sim e^{-\frac{4a}{pr}}/r^{2}$, which vanishes for all available choices of $a$ and $p$. In the Schwarzschild background, the near-horizon behavior of the energy density is given by $\mathcal{H}\sim \left(r-r_s\right)^{(4a)/(pr_s)-1}$. Since $(4a)/(pr_s)>0$ for all choices, the energy density is always at least integrable in a one-sided neighbourhood of the horizon. For the choices  $p=5$, $r_s=1$ and $a=1/4$ used in Fig.~\ref{fig5}, the near-horizon behavior is given by $\mathcal{H}\sim \left(1-r\right)^{1/5}$. }

Solution~\eqref{finalSol} is also valid in the other backgrounds considered in this work (i.e., those in which $\xi$ is unbounded from above and below), with properties similar to those of our previous examples. Moreover, the general orbit~\eqref{orbitnewmodel} can be used to find all other solutions of the first-order equations, as was done for the BNRT model~\cite{fator1}, with the $C=0$ case represented by~\eqref{finalSol} being but one example. In the interest of brevity, and because such an investigation falls outside the overall theme of this work, we shall not discuss all these solutions in detail here. Nevertheless, it is worthwhile to note that they are novel even in the flat $D=1$ case, where few models with exactly integrable first-order equations are known.

\section{Summary and discussion}
	In this work, we have investigated topologically nontrivial solutions in scalar field theories defined on radially symmetric backgrounds of arbitrary dimensions. We start with kink-like solutions found in one-field theories in generalized $D=1$ spacetimes, with and without intrinsic curvature. It is found that the BPS equations of the theories considered can always be put in the standard form $d\phi/d\xi=W_{\phi}$ after a coordinate transformation, where the $\xi$ variable completely encompasses the effect of geometry in the kink-like defects. We then generalize the results to the setting of two-field Lagrangians in spherically symmetric backgrounds with metric tensor of the form~\eqref{metric}, with straightforward extensions possible for cylindrical and similar \qt{radial} backgrounds. A Bogomol'nyi procedure is effected within the symmetric restriction of the theory, obtained by considering solutions $(\phi,\chi)=(\phi(t,r),\chi(t,r))$ that are unchanged by rotations. We are able to obtain \textcolor{black}{radially} stable solitonic solutions despite Derrick's theorem through the use of a potential with explicit radial dependence, similarly to the technique first used in Ref.~\cite{PRL2003} for one-field theories in flat spacetime. The first-order equations whose solutions minimize the energy in a given sector are found, and the orbit equation, from which the general target-space orbits satisfied by physical configurations is deduced. Theories with both canonical and generalized kinetic contributions of the form introduced in Ref.~\cite{liao1} are investigated, and the extra flexibility is used to explore new models where the first-order equations can be fully integrated. In the separable case, the $\phi(r)$ configurations can be viewed effectively as geometrically constrained domain walls with a geometry resulting from a combination of the background space and the separable coupling with the $\chi$ field, whose equation is integrated independently. We have also introduced a generalized model without separability and integrated its orbit equation to solve the problem completely. \textcolor{black}{We have also developed a weak formalism which generalizes the strong equations~\eqref{EL} and~\eqref{FOII}, which need not be strictly defined everywhere. This formalism can also be made more flexible if needed, in order to allow, for example, delta sources or similar distributional discontinuities, which may be needed for some applications and for eventual generalizations of our results.}
	
	It is found that theories derived from a given superpotential share the same orbits regardless of the background geometry, and are related to each other by a one-parameter mapping. The effect of geometry on the solitons is completely determined by the transformation~\eqref{transform}. This allows one to study these systems by treating them as BPS solutions defined in terms of the function $\xi$. If the range of this function is the entire real line, as is the case in three-dimensional flat spacetime, as well as pure de Sitter and Schwarzschild de Sitter backgrounds and the family of conformal metrics represented by~\eqref{conformalmetric}, then solutions can be immediately found by replacing $x$ by $\xi$ in known two-dimensional BPS configurations. On the other hand, two-dimensional solutions with an exponential tail such as those of the $\phi^4$ or BNRT models cannot be translated to backgrounds that give rise to a function $\xi(r)$ is bounded from above or below, as is the case in asymptotically flat spacetimes. Two-dimensional systems which support compacton-like configurations can be mapped into these spaces, since the fields can reach points of $\mathcal{M}$ \textcolor{black}{, or the corresponding limiting boundary values in singular models,} at a finite value of $\xi$.
	
	These results are important in that they provide a simple way to investigate solitons in radially symmetric backgrounds. Scalar fields are ubiquitous in high energy and gravitational physics, and solitonic configurations provide an important window into the role of topology and non-perturbative effects in these settings. BPS solutions in particular have interesting properties and are often found in theories possessing supersymmetry, which is an important ingredient of many quantum gravity candidates. These solutions can also be used as analytical tools that are able to capture the main features of topological solitons, many of which remain valid even outside of BPS saturation. Moreover, scalar solitons may serve as useful prototypical models through which more general features of topological solutions can be investigated. It may also be worthwhile to extend the theories discussed here to allow for more general kinetic terms, which may be useful, for example, in K-essence applications. This is particularly important in the setting of curved spacetimes, since the nonlinear equations that lead to configurations of this kind can become difficult in nontrivial geometries. Further advancements that can build upon our results include applications to brane modeling, where scalar fields are known to be useful for stabilization of inter-brane spacing and to dictate the form of the metric warp factor in a natural way~\cite{Rubakov1983,Goldberger1999, Kehagias2001, Bazeia2004}. Two-field theories are useful for these applications and, in particular, the BNRT model has been successfully used to engender thick branes with rich internal structure~\cite{BazeiaGomes2004}. Our results allow for the generalization of these theories to more general bulk spaces, with curved fifth dimension or even a higher number of extra dimensions, since string theory often requires more than one extra dimension. Some recent works have dealt with the six-dimensional case of this problem, see for example~\cite{6dbraneI, 6dbraneII}. Scalar models in higher dimensions also have an important advantage in the fact that they can couple to other types of particle-like structures that do not exist in two-dimensional spacetimes. This was done, for example, in Ref.~\cite{liao4}, where a one-field scalar theory of the kind introduced in~\cite{PRL2003} was coupled to a Nielsen-Olesen~\cite{Nielsen} type Lagrangian in order to generate an effective impurity-doped vortex system. Because vortices only exist in codimension two, such a coupling would not be possible with a standard Lagrangian. Since, as has been demonstrated in our examples, the soliton size can change freely through variation of the parameter $\xi_{0}$, these solutions can be used to model any kind of \textcolor{black}{radially} stable, symmetric localized structure living in a neighbourhood of a black hole or other compact astrophysical object. Finally, we note that, since this work only concerns static solutions, further investigations are needed in order to understand dynamical aspects of the theory, such as scattering properties.

	\acknowledgments{ This research was funded by the Brazilian agencies Conselho Nacional de Desenvolvimento Cient\'ifico e Tecnol\'ogico (CNPq), grant $\# 151204/2024-1$ (MAL) and Paraiba State Research Foundation (FAPESQ-PB), grant $\# 2783/2023$ (IA).}
	
	
\twocolumngrid

\end{document}